%% file: paper.tex
\documentclass{modelica}
\hypersetup{%
  pdftitle  = {FMI Meets SystemC: A Framework for Cross-Tool Virtual Prototyping},
  pdfauthor = {Nils Bosbach, Meik Schmidt, Lukas Jünger, Matthias Berthold, Rainer Leupers},
  pdfsubject = {16th International Modelica Conference 2025},
  pdfkeywords = {FMI, SystemC, VCML},
  colorlinks,
  linkcolor=black,
  urlcolor=black,
  citecolor=black,
  pdfpagelayout = SinglePage,
  pdfcreator = pdflatex,
  pdfproducer = pdflatex}

\usepackage[printonlyused,nolist]{acronym}
\usepackage[capitalise,nameinlink]{cleveref}
\usepackage{siunitx}
\usepackage{orcidlink}

\makeatletter
\renewcommand*\AC@acs[1]{%
    \expandafter\AC@get\csname fn@#1\endcsname\@firstoftwo{#1}}
\makeatother
\definecolor{rwth}   {RGB}{  0  84 159}
\definecolor{rwth-75}{RGB}{ 64 127 183}
\definecolor{rwth-50}{RGB}{142 186 229}
\definecolor{rwth-25}{RGB}{199 221 242}
\definecolor{rwth-10}{RGB}{232 241 250}

\definecolor{black}   {RGB}{  0   0   0}
\definecolor{black-75}{RGB}{100 101 103}
\definecolor{black-50}{RGB}{156 158 159}
\definecolor{black-25}{RGB}{207 209 210}
\definecolor{black-10}{RGB}{236 237 237}

\definecolor{magenta}   {RGB}{227   0 102}
\definecolor{magenta-75}{RGB}{233  96 136}
\definecolor{magenta-50}{RGB}{241 158 177}
\definecolor{magenta-25}{RGB}{249 210 218}
\definecolor{magenta-10}{RGB}{253 238 240}

\definecolor{yellow}   {RGB}{255 237   0}
\definecolor{yellow-75}{RGB}{255 240  85}
\definecolor{yellow-50}{RGB}{255 245 155}
\definecolor{yellow-25}{RGB}{255 250 209}
\definecolor{yellow-10}{RGB}{255 253 238}

\definecolor{petrol}   {RGB}{  0  97 101}
\definecolor{petrol-75}{RGB}{ 45 127 131}
\definecolor{petrol-50}{RGB}{125 164 167}
\definecolor{petrol-25}{RGB}{191 208 209}
\definecolor{petrol-10}{RGB}{230 236 236}

\definecolor{turkis}   {RGB}{  0 152 161}
\definecolor{turkis-75}{RGB}{  0 177 183}
\definecolor{turkis-50}{RGB}{137 204 207}
\definecolor{turkis-25}{RGB}{202 231 231}
\definecolor{turkis-10}{RGB}{235 246 246}

\definecolor{grun}   {RGB}{ 87 171  39}
\definecolor{grun-75}{RGB}{141 192  96}
\definecolor{grun-50}{RGB}{184 214 152}
\definecolor{grun-25}{RGB}{221 235 206}
\definecolor{grun-10}{RGB}{242 247 236}

\definecolor{maigrun}   {RGB}{189 205   0}
\definecolor{maigrun-75}{RGB}{208 217  92}
\definecolor{maigrun-50}{RGB}{224 230 154}
\definecolor{maigrun-25}{RGB}{240 243 208}
\definecolor{maigrun-10}{RGB}{249 250 237}

\definecolor{orange}   {RGB}{246 168   0}
\definecolor{orange-75}{RGB}{250 190  80}
\definecolor{orange-50}{RGB}{253 212 143}
\definecolor{orange-25}{RGB}{254 234 201}
\definecolor{orange-10}{RGB}{255 247 234}

\definecolor{rot}   {RGB}{204   7  30}
\definecolor{rot-75}{RGB}{216  92  65}
\definecolor{rot-50}{RGB}{230 150 121}
\definecolor{rot-25}{RGB}{243 205 187}
\definecolor{rot-10}{RGB}{250 235 227}

\definecolor{bordeaux}   {RGB}{161  16  53}
\definecolor{bordeaux-75}{RGB}{182  82  86}
\definecolor{bordeaux-50}{RGB}{205 139 135}
\definecolor{bordeaux-25}{RGB}{229 197 192}
\definecolor{bordeaux-10}{RGB}{245 232 229}

\definecolor{lila}   {RGB}{122 111 172}
\definecolor{lila-75}{RGB}{155 145 193}
\definecolor{lila-50}{RGB}{188 181 215}
\definecolor{lila-25}{RGB}{222 218 235}
\definecolor{lila-10}{RGB}{242 240 247}

\definecolor{violett}   {RGB}{ 97  33  88}
\definecolor{violett-75}{RGB}{131  78 117}
\definecolor{violett-50}{RGB}{168 133 158}
\definecolor{violett-25}{RGB}{210 192 205}
\definecolor{violett-10}{RGB}{237 229 234}

\usepackage{xspace}
\usepackage[skip=1pt]{caption}
\usepackage[skip=1pt]{subcaption}
\Crefname{figure}{Figure}{Figure}
\Crefname{table}{Table}{Table}

\usepackage{balance}
\usepackage{lastpage}

\makeatletter
\def\lastreferencepage{\lastpage@lastpage} 
\makeatother
\def\balanceissued{unbalanced}
\renewbibmacro{finentry}{%
    \ifnum\thepage=\lastreferencepage%
        \expandafter\ifx\expandafter\relax\balanceissued\relax%
        \else%
            \balance%
            \gdef\balanceissued{\relax}%
        \fi%
    \fi%
    \finentry%
}

\lstdefinestyle{rwth-style}{
    basicstyle=\ttfamily\footnotesize,
    breakatwhitespace=true,
    breaklines=true,
    captionpos=b,
    commentstyle=\color{grun},
    escapechar=\%,
    floatplacement=htb,
    frame=both,
    keepspaces=true,
    keywordstyle=\color{rwth},
    language=C++,
    numbers=left,
    numbersep=8pt,
    numberstyle=\tiny\color{black-75},
    numberblanklines=true,
    showspaces=false,
    showstringspaces=false,
    showtabs=false,
    stringstyle=\color{bordeaux},
    tabsize=2,
    xleftmargin=2em,
}
\usepackage{tikz}
\usetikzlibrary{
    positioning,
    arrows,
    fit,
    shapes.geometric,
    matrix,
    patterns,
    calc,
    shapes,
}

\usepackage{pgfplots}
\pgfplotsset{compat=1.18}
\pgfplotsset{
    General/.style={
        font=\footnotesize,
        width=\linewidth,
        height=9\baselineskip,
        xtick pos=left,
        xtick align=outside,
        ytick pos=left,
        ytick align=outside,
        ymajorgrids=true,
        grid style=dashed,
        legend cell align={left},
        line width=0.6pt,
        enlarge x limits=0.05,
        mark size=1.5pt,
    },
    BarConfig/.style={
        General,
        ymin=0,
        xtick=data,
        xticklabel style={rotate=30, anchor=east},
        bar width=5pt,
    },
}

\tikzstyle{arrow}     = [thick,->,>=stealth]
\tikzstyle{nodetype}  = [rectangle, rounded corners, minimum height=.4cm, text centered, text width=1.1cm, draw=black, fill=white, font={\footnotesize}, inner sep=1mm, outer sep=0pt]
\tikzstyle{label}     = [draw=none, font={\footnotesize}, inner sep=0pt, outer sep=0pt]
\tikzstyle{startstop} = [rectangle, rounded corners, minimum width=1cm, minimum height=0.7cm,text centered, draw=black, fill=bordeaux-25, font={\footnotesize}]
\tikzstyle{io}        = [trapezium, trapezium left angle=70, trapezium right angle=110, minimum width=2cm, minimum height=0.7cm, text centered, draw=black, fill=rwth-25, font={\footnotesize}]
\tikzstyle{process}   = [rectangle, minimum width=1.5cm, minimum height=0.7cm, text centered, draw=black, fill=orange-25, font={\footnotesize}]
\tikzstyle{decision}  = [diamond, minimum width=2cm, minimum height=0.7cm, text centered, draw=black, fill=grun-25, font={\footnotesize}, aspect=2.5]

\usepackage[pages=some]{background}

\begin{document}

\begin{acronym}[AAPCS64]
    \acro{aapcs64}[AAPCS64]{Procedure Call Standard for the \acl{aarch64}}
    \acro{aarch64}[AArch64]{\acs{arm} 64-bit Architecture}
    \acro{ai}[AI]{Artificial Intelligence}
    \acro{aiba}[AIBA]{An Automated Intra-cycle Behavioral Analysis for SystemC-based design exploration}
    \acro{algol60}[ALGOL~60]{Algorithmic Language 1960}
    \acro{amd}[AMD]{Advanced Micro Devices}
    \acro{ams}[AMS]{Analog/Mixed-Signal}
    \acro{aoa}[AoA]{ARM-on-ARM}
    \acro{api}[API]{Application Programming Interface}
    \acro{arm}[ARM]{Advanced \acs{risc} Machines}
    \acro{asil}[ASIL]{Automotive Safety Integrity Level}
    \acro{ast}[AST]{Abstract Syntax Tree}
    \acro{avp64}[AVP64]{\acs{arm}v8 \acl{vp}}
    \acro{bar}[BAR]{Base Address Register}
    \acro{bb}[BB]{Basic Block}
    \acro{bl}[BL]{Branch-With-Link}
    \acro{bp}[BP]{Breakpoint}
    \acro{can}[CAN]{Controller Area Network}
    \acro{ci}[CI]{Continous Integration}
    \acro{cicd}[CI/CD]{Continuous Integration/Continuous Delivery}
    \acro{clint}[CLINT]{Core-Local Interrupt Controller}
    \acro{cpu}[CPU]{Central Processing Unit}
    \acro{crc}[CRC]{Cyclic Redundancy Check}
    \acro{csv}[CSV]{Character-Separated Values}
    \acro{db}[DB]{Database}
    \acro{dbt}[DBT]{Dynamic Binary Translation}
    \acro{dbms}[DBMS]{\Acl{db} Management System}
    \acro{ddr}[DDR]{Double Data Rate}
    \acro{des}[DES]{Discrete Event Simulation}
    \acro{dla}[DLA]{Deep Learning Accelerator}
    \acro{dma}[DMA]{Direct Memory Access}
    \acro{dmi}[DMI]{Direct Memory Interface}
    \acro{ds}[DS]{Developer Studio}
    \acro{dwarf}[DWARF]{Debugging With Arbitrary Record Formats}
    \acro{ecu}[ECU]{Electronic Control Unit}
    \acro{eda}[EDA]{Electronic Design Automation}
    \acro{eembc}[EEMBC]{Embedded Microprocessor Benchmark Consortium}
    \acro{etrace}[etrace]{Execution Trace}
    \acro{el}[EL]{Exception Level}
    \acro{elf}[ELF]{Executable and Linkable Format}
    \acro{elog}[elog]{Execution Log}
    \acro{esa}[ESA]{European Space Agency}
    \acro{esl}[ESL]{Electronic System Level}
    \acro{fd}[fd]{file descriptor}
    \acro{fig}[FIG]{Fault Injection in glibc}
    \acro{fmi}[FMI]{Functional Mock-up Interface}
    \acro{fmi-ls-bus}[FMI-LS-BUS]{FMI Layered Standard for Network Communication}
    \acro{fmu}[FMU]{Functional Mock-up Unit}
    \acro{fp}[FP]{Frame Pointer}
    \acro{fpga}[FPGA]{Field Programmable Gate Array}
    \acro{fss}[FSS]{Full-System Simulator}
    \acro{fvp}[FVP]{Fixed Virtual Platform}
    \acro{gcc}[GCC]{GNU Compiler Collection}
    \acro{gdb}[GDB]{GNU Debugger}
    \acro{gic}[GIC]{Generic Interrupt Controller}
    \acro{got}[GOT]{Global Offset Table}
    \acro{gpio}[GPIO]{General Purpose Input/Output}
    \acro{gpu}[GPU]{Graphics Processing Unit}
    \acro{gui}[GUI]{Graphical User Interface}
    \acro{hart}[hart]{Hardware Thread}
    \acro{html}[HTML]{HyperText Markup Language}
    \acro{hw}[HW]{Hardware}
    \acro{i2c}[I$^{2}$C]{Inter-Integrated Circuit}
    \acro{ibm}[IBM]{International Business Machines}
    \acro{id}[ID]{identifier}
    \acro{ieee}[IEEE]{Institute of Electrical and Electronics Engineers}
    \acro{io}[I/O]{Input/Output}
    \acro{iommu}[IOMMU]{Input-Output Memory Management Unit}
    \acro{iot}[IoT]{Intenet of Things}
    \acro{iova}[IOVA]{IO Virtual Address}
    \acrodefplural{iova}[IOVAs]{IO Virtual Addresses}
    \acro{ip}[IP]{Internet Protocol}
    \acro{ipa}[IPA]{Intermediate Physical Address}
    \acrodefplural{ipa}[IPAs]{Intermediate Physical Addresses}
    \acro{ir}[IR]{Intermediate Representation}
    \acro{irq}[IRQ]{Interrupt Request}
    \acro{isa}[ISA]{Instruction-Set Architecture}
    \acro{isa-bus}[ISA]{Industry Standard Architecture}
    \acro{iss}[ISS]{Instruction-Set Simulator}
    \acro{jit}[JIT]{Just-In-Time}
    \acro{kvm}[KVM]{Kernel Virtual Machine}
    \acro{l4vecu}[L4 vECU]{Level~4 Virtual Electronic Control Unit}
    \acro{lr}[LR]{Link Register}
    \acro{lt}[LT]{Loosely-Timed}
    \acro{mcdc}[MC/DC]{Modified Condition/Decision Coverage}
    \acro{mcu}[MCU]{Microcontroller Unit}
    \acro{mips}[MIPS]{Million Instructions Per Second}
    \acro{miso}[MISO]{Master Input Slave Output}
    \acro{mmio}[MMIO]{Memory-Mapped Input/Output}
    \acro{mmu}[MMU]{Memory Management Unit}
    \acro{mnist}[MNIST]{Modified National Institute of Standards and Technology}
    \acro{mosi}[MOSI]{Master Output Slave Input}
    \acro{msi}[MSI]{Message Signalled Interrupt}
    \acro{nas}[NAS]{\acs{nasa} Advanced Supercomputing}
    \acro{nasa}[NASA]{National Space Agency}
    \acro{nistt}[NISTT]{A Non-Intrusive SystemC-TLM 2.0 Tracing Tool}
    \acro{nn}[NN]{Neural Network}
    \acro{nop}[NOP]{No Operation}
    \acro{npb}[NPB]{\acs{nas} Parallel Benchmarks}
    \acro{nvdla}[NVDLA]{NVIDIA \acl*{dla}}
    \acro{openmp}[OpenMP]{Open Multi-Processing}
    \acro{os}[OS]{Operating System}
    \acro{pa}[PA]{Physical Address}
    \acrodefplural{pa}[PAs]{Physical Addresses}
    \acro{pc}[PC]{Program Counter}
    \acro{pccts}[PCCTS]{Purdue Compiler Construction Tool Set}
    \acro{pci}[PCI]{Peripheral Component Interconnect}
    \acro{pcie}[PCIe]{PCI Express}
    \acro{pid}[PID]{process ID}
    \acro{plic}[PLIC]{Platform-Level Interrupt Controller}
    \acro{plt}[PLT]{Procedure Linkage Table}
    \acro{pmu}[PMU]{Performance Monitoring Unit}
    \acro{pthread}[p\-thread]{\acs{posix} Thread}
    \acro{posix}[POSIX]{Portable Operating System Interface}
    \acro{pwm}[PWM]{Pulse Width Modulation}
    \acro{qemu}[QEMU]{Quick Emulator}
    \acro{ram}[RAM]{Random-Access Memory}
    \acro{risc}[RISC]{Reduced Instruction Set Computer}
    \acro{rsp}[RSP]{Remote Serial Protocol}
    \acro{rtc}[RTC]{Real-Time Clock}
    \acro{rtf}[RTF]{Real-Time Factor}
    \acro{rtl}[RTL]{Register-Transfer Level}
    \acro{rtos}[RTOS]{real-time operating system}
    \acro{sata}[SATA]{Serial AT Attachment}
    \acro{sd}[SD]{Secure Digital}
    \acro{sdhci}[SDHCI]{\acs{sd} Host Controller Interface}
    \acro{simd}[SIMD]{Single Instruction, Multiple Data}
    \acro{smmu}[SMMU]{System Memory Management Unit}
    \acro{soc}[SoC]{System-on-a-Chip}
    \acrodefplural{soc}[SoCs]{Systems-on-Chips}
    \acro{sp}[SP]{Stack Pointer}
    \acro{spi}[SPI]{Serial Peripheral Interface}
    \acro{sql}[SQL]{Structured Query Language}
    \acro{ssd}[SSD]{Solid State Drive}
    \acro{sw}[SW]{Software}
    \acro{tb}[TB]{Translation Block}
    \acro{tcg}[TCG]{Tiny Code Generator}
    \acro{tcp}[TCP]{Transmission Control Protocol}
    \acro{td}[TD]{Temporal Decoupling}
    \acro{tfl}[TFLite]{TensorFlow Lite}
    \acro{tlm}[TLM]{Transaction-Level Modeling}
    \acro{tpu}[TPU]{Tensor Processing Unit}
    \acro{uart}[UART]{Universal Asynchronous Receiver/Transmitter}
    \acro{unix}[UNIX]{Uniplexed Information and Computing Service}
    \acro{va}[VA]{Virtual Address}
    \acrodefplural{va}[VAs]{Virtual Addresses}
    \acro{vcd}[VCD]{Value Change Dump}
    \acro{vcml}[VCML]{Virtual Components Modeling Library}
    \acro{vcpu}[vCPU]{virtual CPU}
    \acro{vfio}[VFIO]{Virtual Function I/O}
    \acro{vhe}[VHE]{Virtualization Host Extensions}
    \acro{viper}[VIPER]{Virtual Platform Explorer}
    \acro{vsp}[VSP]{VCML Session Protocol}
    \acro{vpci}[vPCI]{virtual PCI}
    \acro{vm}[VM]{Virtual Machine}
    \acro{vp}[VP]{Virtual Platform}
    \acro{vt-d}[VT-d]{Virtualization Technology for Directed I/O}
    \acro{wfi}[WFI]{Wait For Interrupt}
\end{acronym}

\thispagestyle{empty}

\title{FMI Meets SystemC: A Framework for\\Cross-Tool Virtual Prototyping}
\author[1]{Nils Bosbach\orcidlink{0000-0002-2284-949X}}
\author[1]{Meik Schmidt}
\author[2]{Lukas Jünger\orcidlink{0000-0001-9149-1690}}
\author[3]{Matthias Berthold}
\author[1]{Rainer Leupers\orcidlink{0000-0002-6735-3033}}
\affil[1]{RWTH Aachen University, Aachen, Germany, {\small\texttt{\{bosbach,schmidtm,leupers\}@ice.rwth-aachen.de}}}
\affil[2]{MachineWare GmbH, Aachen, Germany, {\small\texttt{lukas@mwa.re}}}
\affil[3]{tracetronic GmbH, Dresden, Germany, {\small\texttt{Matthias.Berthold@tracetronic.de}}}

\newcommand{\ecutest}{ecu.test\xspace}

\newcommand\copyrighttext{}

\newcommand\copyrightnotice{%
    \backgroundsetup{opacity=1, scale=1, angle=0, contents={
            \color{black}%
            \begin{tikzpicture}[remember picture,overlay]%
                \node[anchor=north,yshift=-10pt,text=gray] at (current page.north) {\shortstack[c]{\large PREPRINT - accepted by the \textit{16th International Modelica and FMI Conference 2025}\\DOI: \href{https://doi.org/10.3384/ecp218545}{\color{gray}{10.3384/ecp218545}}}};
            \end{tikzpicture}%
        }%
    }%
    \BgThispage%
}

\maketitle\thispagestyle{empty} 
\copyrightnotice

\acused{cpu}
\acused{tlm}
\abstract{%
As systems become more complex, the demand for thorough testing and virtual prototyping grows.
To simulate whole systems, multiple tools are usually needed to cover different parts.
These parts include the hardware of a system and the environment with which the system interacts.
The \ac{fmi} standard for co-simulation can be used to connect these tools.

The control part of modern systems is usually a computing unit, such as a \ac{soc} or \ac{mcu}, which executes software from a connected memory and interacts with peripherals.
To develop software without requiring access to physical hardware, full-system simulators, the so-called \acp{vp}, are commonly used.
The \acs{ieee}-standardized framework for \ac{vp} development is SystemC \ac{tlm}.
SystemC provides interfaces and concepts that enable modular design and model exchange.
However, SystemC lacks native \ac{fmi} support, which limits the integration into broader co-simulation environments.

This paper presents a novel framework to control and interact with SystemC-based \acp{vp} using the \ac{fmi}.
We present a case study showing how a simulated temperature sensor in a SystemC simulation can obtain temperature values from an external tool via \ac{fmi}.
This approach allows the unmodified target software to run on the \ac{vp} and receive realistic environmental input data such as temperature, velocity, or acceleration values from other tools.
Thus, extensive software testing and verification is enabled.
By having tests ready and the software pre-tested using a \ac{vp} once the physical hardware is available, certifications like ISO~26262 can be done earlier.
}


\acresetall
\acused{cpu}
\acused{tcp}
\acused{ip}
\acused{gdb}
\acused{spi}
\acused{can}
\acused{i2c}

\section{Introduction}
\label{sec:introduction}

In today's rapidly evolving technology landscape, systems are becoming more complex, requiring advanced development cycles and rigorous testing methodologies.
As industries, such as automotive, tend to solve more problems in software rather than hardware, early and intensive software testing is paramount \cite{ondrej_burkacky_rethinking_2018}.
To start software development and testing while the hardware is still under design, \acp{vp}, also known as \textit{\acp{l4vecu}}, can be used.
It has been shown that the earlier bugs are found, the lower the cost of rectifying them.
Fixing a bug during the development phase can be up to \num{100} times cheaper than fixing it while the product is already in use \cite{boehm_software_1976}.
\acp{vp} are the key enabler of early software testing.
By simulating a \ac{soc}, the unmodified target\footnote{\textit{target}: architecture that is simulated; \textit{host}: architecture that runs the simulation} software can be developed and tested without requiring access to the physical hardware.

The \acs{ieee}-standardized framework for \ac{vp} design is SystemC with its \ac{tlm} extension \cite{systemc-2023}.
While \acp{vp} are well-suited tools for software development and verification, they reach their limits when it comes to realistic testing of different scenarios.
Because embedded systems, such as \acp{ecu}, interact with their physical environment by reading sensor values and controlling devices, it is not enough to simulate only the system itself.
In addition, environmental effects must also be considered and simulated.

To simulate physical systems and define complex test cases, there are many tools available.
In this paper, we show how a SystemC-based simulation can be connected to those tools using the \ac{fmi} standard \cite{modelica_association_fmi_2024}.
The \ac{fmi} co-simulation standard allows to synchronize the simulation time of multiple simulations and to exchange data.

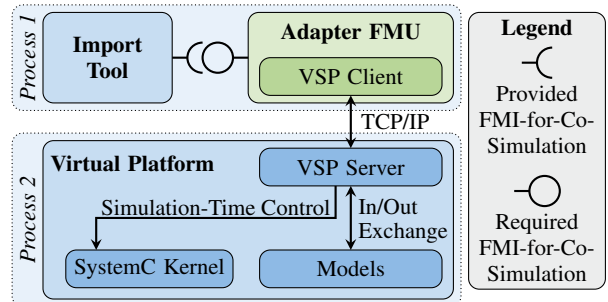
\begin{figure}
  \centering
  \input{tikz/schema.tikz}
  \caption{SystemC-\acs*{fmu} adapter approach.}
  \label{fig:approach}
  \vspace{-\baselineskip}
\end{figure}

\Cref{fig:approach} shows the integration of our approach in a co-simulation.
We developed an \ac{fmu} that can be imported by any \ac{fmi}-compatible tool to control a \ac{vp}.
This \ac{fmu} exposes the specified output of \ac{vp} and accepts the inputs that are forwarded to \ac{vp}.
The \ac{vp} runs in a second process, either on the same host as the \ac{fmi} import tool, or on a different one.
The connection between the \ac{fmu} and the \ac{vp} is \ac{tcp}/\ac{ip} based.

We present a case study where we show how an open-source \ac{vp} can be connected to tracetronic's \textit{\ecutest} tool using \ac{fmi}.
\ecutest allows defining complex test scenario where environmental data can be specified and responses of the tested system are compared against expectations.

In the end, we propose a software design flow that uses our integration for early testing and verification.
This design flow helps to shorten the time to reach specifications like ISO~26262 \cite{iso-26262}.

\section{Background and Related Work}
\label{sec:background}
This section provides an overview of the related work that forms the foundation of this study.
\Cref{sec:background:fmi} explains the \ac{fmi} standard and the features used in this work.
\Cref{sec:background:virtual-prototyping} provides an overview of virtual prototyping, including background information on the industry-standard framework SystemC and the \ac{vcml}, which builds on top of SystemC.
In \cref{sec:background:related-work}, an overview of existing approaches to integrate SystemC modules into an \ac{fmi} co-simulation is given.

\subsection{\acf{fmi}}
\label{sec:background:fmi}

The \acf{fmi} is an open standard designed to enable seamless integration and interoperability of simulation models from various tools and vendors \cite{fmi-3.0.2}.
Initially introduced by the Modelica Association in 2010, \ac{fmi} has become a widely adopted standard for modular and tool-agnostic simulation workflows and is supported by more than \num{200} tools.

Simulation models are encapsulated into so-called \acp{fmu}.
This encapsulation process is performed by \textit{export tools}, which are typically provided by simulation-software vendors.
\acp{fmu} can then be imported and utilized by \textit{import tools}, which coordinate the simulation and facilitate data exchange between multiple \acp{fmu}.

\begin{figure}[!t]
  \centering
  \begin{subfigure}[c]{0.48\linewidth}
    \centering
    \input{tikz/fmi_me.tikz}
    \caption{Model Exchange.}
    \label{fig:fmi-modes:me}
    \vspace{1em}
  \end{subfigure}
  \begin{subfigure}[c]{0.48\linewidth}
    \centering
    \input{tikz/fmi_cs.tikz}
    \caption{Co-Simulation.}
    \label{fig:fmi-modes:cs}
    \vspace{1em}
  \end{subfigure}
  \caption{\acs{fmi} types.}
  \label{fig:fmi-modes}
  \vspace{-\baselineskip}
\end{figure}
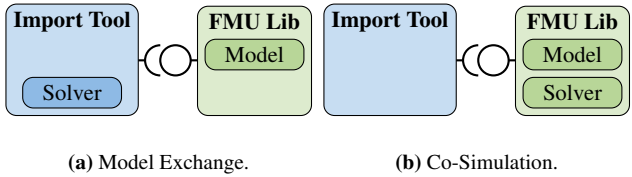

The original standard supports two interface types, \textit{Model Exchange} \cite{fmi-me-1.0} and \textit{Co-Simulation} \cite{fmi-cs-1.0}, as illustrated in \cref{fig:fmi-modes}.
As physical systems are often described by differential equations, a numerical solver is needed to solve and evaluate those equations.
The difference between the \ac{fmu} interface types is primarily in the placement of the numerical solver.
In the Model-Exchange use-case, the solver is not part of the \ac{fmu} and therefore needs to be implemented in the import tool.

With the introduction of FMI 3.0 \cite{fmi-3.0.2}, a third interface type, \textit{Scheduled Execution}, was added.
For this work, we only use the Co-Simulation interface, so the other ones are not further discussed.
We demonstrate how a full \ac{vp} can be encapsulated into an \ac{fmu}.
Because \acp{vp} usually operate at a much higher level of abstraction than differential equations, we do not require a numerical solver.
Consequently, Co-Simulation is the preferred approach.

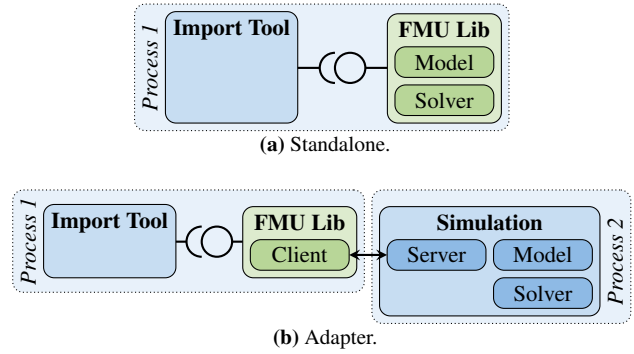
\begin{figure}[!t]
  \centering
  \begin{subfigure}[c]{\linewidth}
    \centering
    \input{tikz/fmi_cs_standalone.tikz}
    \caption{Standalone.}
    \vspace{1em}
    \label{fig:fmi-approaches:standalone}
  \end{subfigure}\\

  \begin{subfigure}[c]{\linewidth}
    \centering
    \input{tikz/fmi_cs_adapter.tikz}
    \caption{Adapter.}
    \vspace{1em}
    \label{fig:fmi-approaches:adapter}
  \end{subfigure}
  \caption{\acs{fmi} Co-Simulation scenarios.}
  \label{fig:fmi-approaches}
  \vspace{-\baselineskip}
\end{figure}

When the Co-Simulation interface is used, two scenarios are possible, a \textit{standalone} and an \textit{adapter} scenario \cite{blochwitz_functional_2011}.
Both scenarios are visualized in \cref{fig:fmi-approaches}.
The \ac{fmu} contains a shared library (\textit{FMU Lib}) that is loaded by the import tool.
It implements standardized \ac{fmi} functions, the import tool calls.
In the standalone scenario, as shown in \cref{fig:fmi-approaches:standalone}, the shared library contains the full simulation model.
The simulation model can be directly executed in the process of the import tool.

In the adapter scenario, illustrated in \cref{fig:fmi-approaches:adapter}, the shared library acts as an interface to another process that executes the simulation.
This connection can, e.g.,  be established using network-based protocols, such as \ac{tcp}/\ac{ip}.
One key advantage of this approach is the ability to run the import tool and simulation on different machines, which can offer potential performance benefits by distributing computational load.
Additionally, this configuration allows the simulation and import tool to run on different \acp{os} or even different hardware architectures, such as x86 and Arm.

\begin{lstlisting}[language=xml, float=*, caption={Model-description file.}, label=lst:mod-desc, abovecaptionskip=1pt, belowcaptionskip=-2em]
<fmiModelDescription ... fmiVersion="3.0" modelName="myVP" ...>
    <CoSimulation modelIdentifier="myVP" needsExecutionTool="true" canHandleVariableCommunicationStepSize="true" ... />
    <DefaultExperiment startTime="3" stopTime="5" stepSize="0.01"/>
    <ModelVariables>
        <Float64 name="time" valueReference="0" causality="independent" variability="continuous" />
        <Float32 name="system.max31855.temp" valueReference="1" causality="input" variability="continuous" start="10.0" />
        <UInt32 name="system.gpio.data" valueReference="2" causality="output" variability="discrete" />
    </ModelVariables>
    <ModelStructure>
        <InitialUnknown valueReference="1"/>
        <Output valueReference="2"/>
    </ModelStructure>
     <Annotations>
        <Annotation type="VCML">
            <VP host="localhost" port="8888"  executable="resources/vp" ... />
        </Annotation>
      </Annotations>
</fmiModelDescription>
\end{lstlisting}

\subsection{Virtual Prototyping}
\label{sec:background:virtual-prototyping}

Virtual prototyping is a technique in which electronic systems are modeled entirely in software.
This work focuses on \acp{vp}, virtual prototypes designed for software development for the simulated target platform.
\acp{vp} can simulate the execution of an unmodified target-software stack.
In the automotive domain, they are referred to as \textit{\acp{l4vecu}}.

The industry-standard framework for \ac{vp} development is SystemC with its \ac{tlm} extension \cite{systemc-2023}.
SystemC standardizes the interface and functionality of a C++ library that features interfaces, data types, and a \ac{des}-based scheduler to simulate parallelism.
A simulation usually contains multiple modules that can be connected via \ac{tlm} sockets.
\ac{tlm} abstracts the underlying communication protocol and offers low-overhead data exchange.
The standardized interfaces guarantee seamless integration of different models into a simulation.

On top of SystemC, modeling libraries provide additional functionalities.
One open-source modeling library is \ac{vcml} \cite{machineware_machineware-gmbhvcml_2024}.
\ac{vcml} extends SystemC with essential components, such as register models, prebuilt hardware components, and specialized \ac{tlm}-based communication protocols.
A key feature used in this work is the \ac{vsp}.
\ac{vsp} is a remote-control protocol that is based on \ac{gdb}'s \ac{rsp} protocol \cite{free_software_foundation_inc_remote_2024}.
\ac{vcml} includes a \ac{vsp} server module, which allows clients to connect via \ac{tcp} to control the simulation.

Another crucial \ac{vcml} feature used in this work is \textit{properties}, which encapsulate configuration variables within modules.
These properties can be set at \ac{vp} launch and accessed via the \ac{vsp} protocol, facilitating dynamic reconfiguration.

\subsection{Related Work}
\label{sec:background:related-work}

Previously, various publications have described methods for integrating SystemC-based simulations into \ac{fmi} co-simulations.
Most of these approaches are based on the \ac{fmi}~2 standard and involve tight coupling between the \ac{fmi} interface and the \ac{vp}.

One approach integrates an \ac{fmi} import tool directly into the \ac{vp} to co-simulate SystemC models with \acp{fmu} that provide environmental models \cite{safar_virtual_2018}.
This approach means that the \ac{vp} cannot be used with other import tools, e.g., for setting up test scenarios, as described in \cref{sec:case-study}.
Furthermore, the \ac{vp} is no longer standalone, so co-simulated \acp{fmu} are always required.

For detailed \ac{rtl}-like SystemC simulations, a different approach can be used to expose SystemC ports to \ac{fmi} \cite{centomo_using_2016}.
This approach also requires modifications to the \ac{vp}, which prevents a standalone simulation.
In our work, we focus on SystemC-\ac{tlm}-based \acp{vp} that use a higher abstraction level to achieve increased performance.

For SystemC \ac{ams} simulations, wrapper classes can be created to wrap models and connect the needed inputs and outputs to an \ac{fmi} integration \cite{krammer_standard_2015}.
This kind of simulation also targets more detailed abstraction levels and requires modifications to the simulation.

In another approach, \textit{\ac{fmi} variables} are introduced, that can be bound to different protocols like \ac{can}, or \ac{i2c} \cite{saidi_fast_2019}.
When a variable changes, different actions can be carried out.
For example, a \ac{can} frame carrying the updated variable as payload can be injected onto the bus.
This approach is useful for simple sensors that use common protocols.
An advantage is that the sensor itself does not need to be part of the \ac{vp} but can be completely used from the \ac{fmi} integration.
This approach also does not allow for standalone use of the \ac{vp}.

In our work, we present an approach that does not require \ac{vp} modifications and therefore no source-code access.
The requirement is that the\ac{vp} is based on the used modeling library \ac{vcml} and the \ac{vsp} server is not disabled.
The \ac{vp} can then be used without the \ac{fmi} integration when environmental input are not needed.
When environmental inputs become important, the \ac{vp} can be integrated into an \ac{fmi} co-simulation without any changes by using our standalone \ac{fmi}-adapter \ac{fmu}.

\section{Approach}
\label{sec:approach}

In this section, we present our approach.
First, \cref{sec:approach:scenario} describes the application scenario.
Then, \cref{sec:approach:fmu-creation} explains how we integrate a \ac{vcml}-based \ac{vp} into an \ac{fmu}.
In the end, \cref{sec:approach:working-principle} provides further details on the \ac{fmu} implementation and the integration process.

\subsection{Application Scenario}
\label{sec:approach:scenario}

Imagine a \ac{vp} that models an \ac{ecu} that consists of various peripherals including sensing devices for measuring environmental conditions.
Such a \ac{vp} is crucial for target software development and verification but traditionally lacks dynamic environmental modeling capabilities.
For instance, a virtual model of a temperature sensor simulates all hardware interfaces other components can access but typically returns constant values during standalone \ac{vp} execution, limiting its ability to test software responses to varying conditions.
Directly integrating environmental simulations, such as temperature models, into device models like sensors would have multiple drawbacks:

\begin{itemize}
  \item \textbf{Simplicity}: The model should be kept simple to have a high performance.
    Complex algorithms to model the environment would unnecessarily slow down the model and increase its complexity when the environmental simulation is not needed.

  \item \textbf{Genericity}: The model should only reflect the behavior of the hardware to use it in various scenarios.
    An environmental simulation model heavily depends on the scenario and not the component itself.

  \item \textbf{Specialization of Tools}: There are well-established tools to create environmental models, e.g., based on physical equations or test cases.
\end{itemize}

Because of those reasons, the returned temperature value from our model is read from a \ac{vcml} property (see \cref{sec:background:virtual-prototyping}).
This means that in a standalone simulation, the temperature value is set at startup and remains constant throughout execution.
While this setup allows for full software stack execution including sensor interaction, it does not support testing of how the software reacts to changing environmental conditions.

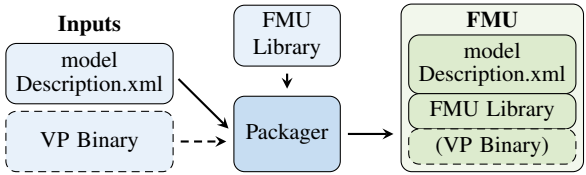
\begin{figure}[!t]
  \centering
  \input{tikz/creation.tikz}
  \vspace{1em}
  \caption{\acs{fmu} creation.}
  \label{fig:creation}
\end{figure}

To overcome this limitation, our \ac{fmi} integration enables co-simulation of the \ac{vp} together with external tools that dynamically provide environmental data.
This approach allows for realistic scenario testing and a more comprehensive evaluation of the system's behavior using dedicated tools to solve various tasks.

\subsection{\acs{fmu} Creation}
\label{sec:approach:fmu-creation}

To create an \ac{fmu} from a \ac{vcml}-based \ac{vp}, our \textit{Packager} tool can be used.
\Cref{fig:creation} illustrates the creation process.
The resulting \ac{fmu} includes our \ac{vp}-independent \ac{fmu} library and a model-description file, which defines its properties.
\Cref{sec:approach:working-principle} explains the working principle of the \ac{fmu} library in detail.
A simplified example of a model-description file is shown in \cref{lst:mod-desc}.

The \lstinline[language=xml]{ModelVariables} element lists the \ac{vp}'s inputs and outputs that are accessible via \ac{fmi}.
Each variable's name corresponds to the property name within the module hierarchy of the \ac{vp}.
For example, the \lstinline{system.max31855.temp} property defines the property \lstinline{temp} of the \ac{spi} temperature sensor \lstinline{max31855}, which is a submodule of the \lstinline{system} module.
It is sufficient to list the hierarchical name of the property in the model-description file.
There are no changes in the \ac{vp} needed to expose the property.
The \ac{vsp} server can clearly identify and access the properties by their hierarchical names.
Additionally, data types are specified, and it is indicated whether the property functions as an input or output.
For inputs, a start value can be assigned.
During simulation, \ac{fmi} uses the \lstinline[language=xml]{valueReference} numbers when updating the variables.

The \lstinline[language=xml]{Annotations} element defines \ac{vp}-specific properties using the available attributes listed in \cref{tab:attributes}.
The \ac{vp} binary can be embedded within the \ac{fmu} and launched by the \ac{fmu}.
In this case, the \lstinline{executable} attribute points to the relative path of the \ac{vp} binary within the \ac{fmu} (e.g., \texttt{resources/my\_vp}).
Additionally, the \lstinline{host} is set to \textit{localhost}, and the \lstinline{port} specifies the \ac{tcp} port the \ac{vp}'s \ac{vsp} server is listening.

\begin{table}[b!]
  \centering
  \vspace{-1em}
  \caption{Attributes of the VCML annotation node.}
  \label{tab:attributes}
  \begin{tabular}{cp{5cm}}
    \toprule
    \multicolumn{1}{c}{\textbf{Attribute}} & \multicolumn{1}{c}{\textbf{Meaning}}                           \\ \midrule
    \texttt{executable}                    & Path of the \acs{vp} executable                                \\
    \texttt{args}                          & Args that are passed to the \acs{vp} executable during startup \\
    \texttt{host}                          & IP address or hostname of the machine that executes the VP     \\
    \texttt{port}                          & Listening port of the \acs{vp} server                          \\ \bottomrule
  \end{tabular}
\end{table}

If the \ac{vp} is not embedded in the \ac{fmu}, e.g., because it is manually launched on a separate machine, only the \lstinline{host} and \lstinline{port} attributes need to be configured.
The \lstinline{host} attribute defines the hostname of the machine that executes the \ac{vp}.
This machine needs to be reachable by the import tool via a \ac{tcp} connection.

\subsection{Working Principle}
\label{sec:approach:working-principle}

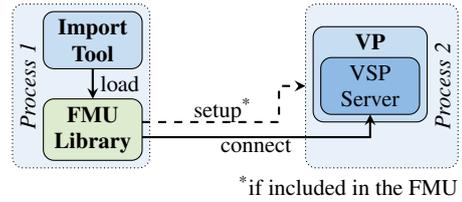
\begin{figure}[!t]
  \centering
  \input{tikz/running_mode.tikz}
  \vspace{1em}
  \caption{Working principle.}
  \label{fig:working-prinicble}
\end{figure}

An exported \ac{fmu} can be loaded by an \ac{fmi}-3-compatible import tool.
\Cref{fig:working-prinicble} illustrates the workflow.

The import tool unpacks the \ac{fmu} and reads the model-description file.
It then loads the \ac{fmu} library (\textit{.dll} or \textit{.so}).
Afterward, the import tool can call the standardized \ac{fmi} functions that are implemented by the \ac{fmu} library.
If the \ac{vp} binary is packed into the \ac{fmu}, the \ac{fmu} library starts the \ac{vp} in a second process.
The library establishes a \ac{tcp}-based connection to the \ac{vsp} server of the \ac{vp} and uses this connection to control the simulation.

\begin{figure}[!t]
  \centering
  \input{tikz/call_seq.tikz}
  \vspace{1em}
  \caption{\acs{fmi} functions implemented by our \acs{fmu}.}
  \label{fig:call-seq}
  \vspace{-1em}
\end{figure}
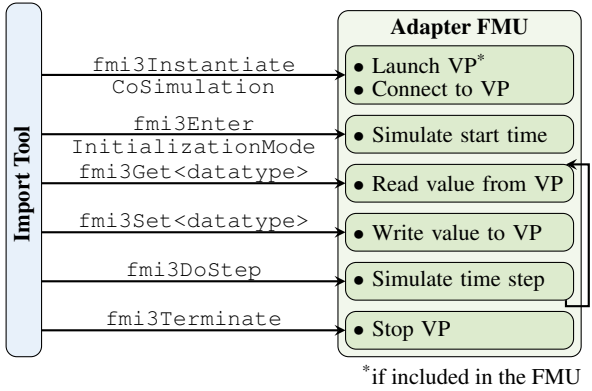

\Cref{fig:call-seq} shows the \ac{fmi} functions our \ac{fmu} implementation.
The tasks that are executed by those functions are explained in the following in more detail.

First, the \lstinline[language=C]{fmi3InstantiateCoSimulation} function is usually called.
If the \ac{vp} is part of the \ac{fmu}, a second process is started to run the \ac{vp}.
Parameters can be passed to the \ac{vp} via command-line arguments.
If the \ac{vp} is not part of the \ac{fmu}, it needs to be started manually.
This can be done either on the same machine or on a different one that is located in the same network as the machine that runs the \ac{fmu}.
In this case, the corresponding hostname of the machine that runs the \ac{vp} needs to be set in the \texttt{VCML} node of the model description's \lstinline[language=xml]{Attribute} node.
Once the \ac{vp} is started, the \ac{vsp} server listens on the \ac{tcp} port for a connection and suspends the \ac{vp} until the \ac{fmu} is connected.
The \ac{fmu} connects the \ac{vsp} server of the \ac{vp}.

When the \lstinline[language=C]{fmi3EnterInitializationMode} function is called by the import tool, a command is sent to the \ac{vsp} server to simulate until the virtual simulation time of the SystemC simulation reaches the specified \lstinline[language=xml]{startTime}.
The start time parameter of the \ac{fmu} can be used to, e.g., skip the boot process of the \ac{os} running on the system.
If the start time is set to \num{0}, our \ac{fmu} does nothing in this step.
After the \lstinline[language=C]{fmi3EnterInitializationMode} function call, the \ac{vp} is ready for simulation.

The import tool can read model variables that have been declared as \textit{output} in the model description by calling the \lstinline[language=C]{fmi3Get<datatype>} functions.
For example, \lstinline[language=C]{fmi3GetFloat64} is used when the variable is declared as \textit{Float64} in the model description.
The variable is referenced by its \lstinline[language=xml]{valueReference} property.
The \ac{fmu} translates the \lstinline[language=xml]{valueReference} into the \lstinline[language=xml]{name} using the model description.
This name corresponds to the hierarchical name of the property in the \ac{vp} (see \cref{sec:approach:fmu-creation}).
This hierarchical name can be directly used by the \ac{vsp} server to identify, read, and write the property value within the simulation.
All available properties of the simulation can be used as \ac{fmu} inputs and outputs by simply specifying their hierarchical name in model description.

To update an \ac{fmu} input, the \lstinline[language=C]{fmi3Set<datatype>} function can be used in the same way.
Again, the variables are referenced by their \lstinline[language=xml]{valueReference} property.
The updated input values are directly sent to the \ac{vp} to update the corresponding model properties.

After variables have been updated, the \lstinline[language=C]{fmi3DoStep} function can be called to simulate a time step.
The amount of simulation time that is simulated on a call is usually defined by the \lstinline[language=xml]{stepSize} attribute in the model description (see \cref{lst:mod-desc}).
Our \ac{fmu} sends a command to the connected \ac{vp} to simulate the step.
The \ac{vp}'s SystemC scheduler simulates events until the virtual simulation time is increased by the specified step.
This allows the virtual \ac{cpu} model to execute target software, and interact with other simulated peripherals.
The import tool repeatedly calls the \lstinline[language=C]{fmi3Get<datatype>}, \lstinline[language=C]{fmi3Set<datatype>}, and \lstinline[language=C]{fmi3DoStep} functions as long as the test or the simulation is running.

To end the simulation, the \lstinline[language=C]{fmi3Terminate} function is called, which stops the \ac{vp} and shuts down the simulation.

\section{Case Study}
\label{sec:case-study}

To demonstrate the effectiveness of our \ac{fmi} integration approach for SystemC-based \acp{vp}, we present a case study of a software-based Schmitt trigger application with a temperature input.
We show how the different components of the system can be efficiently modeled and integrated into an \ac{fmi}-based co-simulation using our developed framework.
The full system consists of three parts:

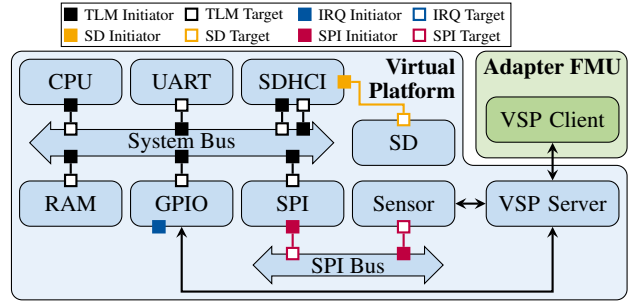
\begin{figure}[!t]
  \centering
  \input{tikz/vp.tikz}
  \vspace{1em}
  \caption{\acs*{avp64}-based \acs{ecu} with connected \acs*{fmi} adapter.}
  \label{fig:vp}
  \vspace{-1em}
\end{figure}

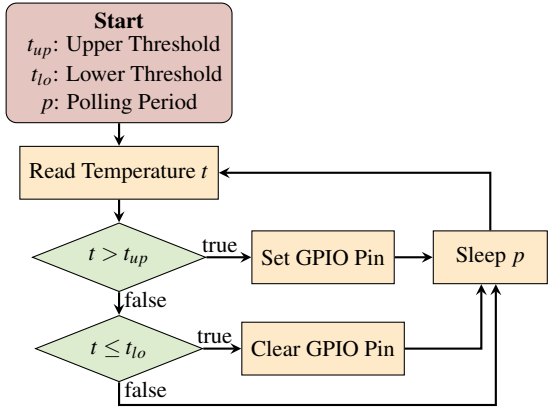
\begin{figure}[!t]
  \centering
  \input{tikz/algorithm.tikz}
  \vspace{1em}
  \caption{Schmitt-trigger algorithm executed by the \ac{vp}.}
  \label{fig:alg}
\end{figure}

\textbf{\ac{ecu}}: The \ac{ecu} consists of an ARMv8 processor, memory, and peripherals.
The components and their connections are shown in \cref{fig:vp}.
A \textit{MAX31855} \cite{maxim_integrated_max31855_2015} temperature sensor (denoted as \textit{Sensor} in \cref{fig:vp}) can measure the ambient temperature.
The sensor is connected via an \ac{spi} interface to the \ac{spi} controller, which is accessible by the \ac{cpu}.
Additionally, the system features a \ac{gpio} controller that allows the \ac{cpu} to control external signals.
To simulate the \ac{ecu} hardware, the open-source, SystemC-\ac{tlm}-based \ac{avp64} \cite{junger_fast_2019,junger_armv8_2023} is used.
The \ac{vp} is capable of executing the full target-software stack.

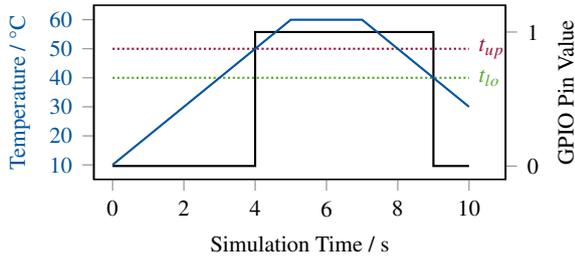
\begin{figure}[!t]
  \centering
  \input{tikz/expected_behavior.tikz}
  \vspace{1em}
  \caption{Temperature input and expected \acs{gpio}-pin output for $t_{lo}=\SI{40}{\degreeCelsius}$ and $t_{up}=\SI{50}{\degreeCelsius}$.}
  \label{fig:data}
  \vspace{-1em}
\end{figure}

\textbf{Software Application}: The \ac{ecu} boots a \textit{Buildroot}-based Linux \ac{os} \cite{buildroot_making_2024}.
Upon completion of the boot process, a simple Schmitt-trigger application is automatically launched.
The Schmitt-trigger application implements the algorithm depicted in \cref{fig:alg}.
This algorithm monitors the temperature input of the sensor and sets or clears a \ac{gpio} pin based on the predefined threshold values $t_{lo}$ and $t_{up}$.
When the temperature exceeds $t_{up}$, the \ac{gpio} pin is set.
When it falls below $t_{lo}$, the \ac{gpio} pin is cleared.
The application periodically polls the temperature sensor at a configurable interval $p$.
An example of a temperature curve and the expected output for $t_{lo}=\SI{40}{\degreeCelsius}$ and $t_{up}=\SI{50}{\degreeCelsius}$ is shown in \cref{fig:data}.

\begin{figure}[t]
  \centering
  \input{tikz/fmu_schema.tikz}
  \vspace{1em}
  \caption{FMU with inputs and outputs.}
  \label{fig:fmu_schema}
  \vspace{-1em}
\end{figure}
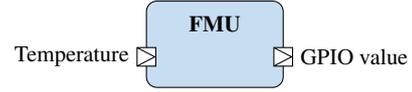

\begin{figure*}[!t]
  \centering
  \includegraphics[width=\linewidth]{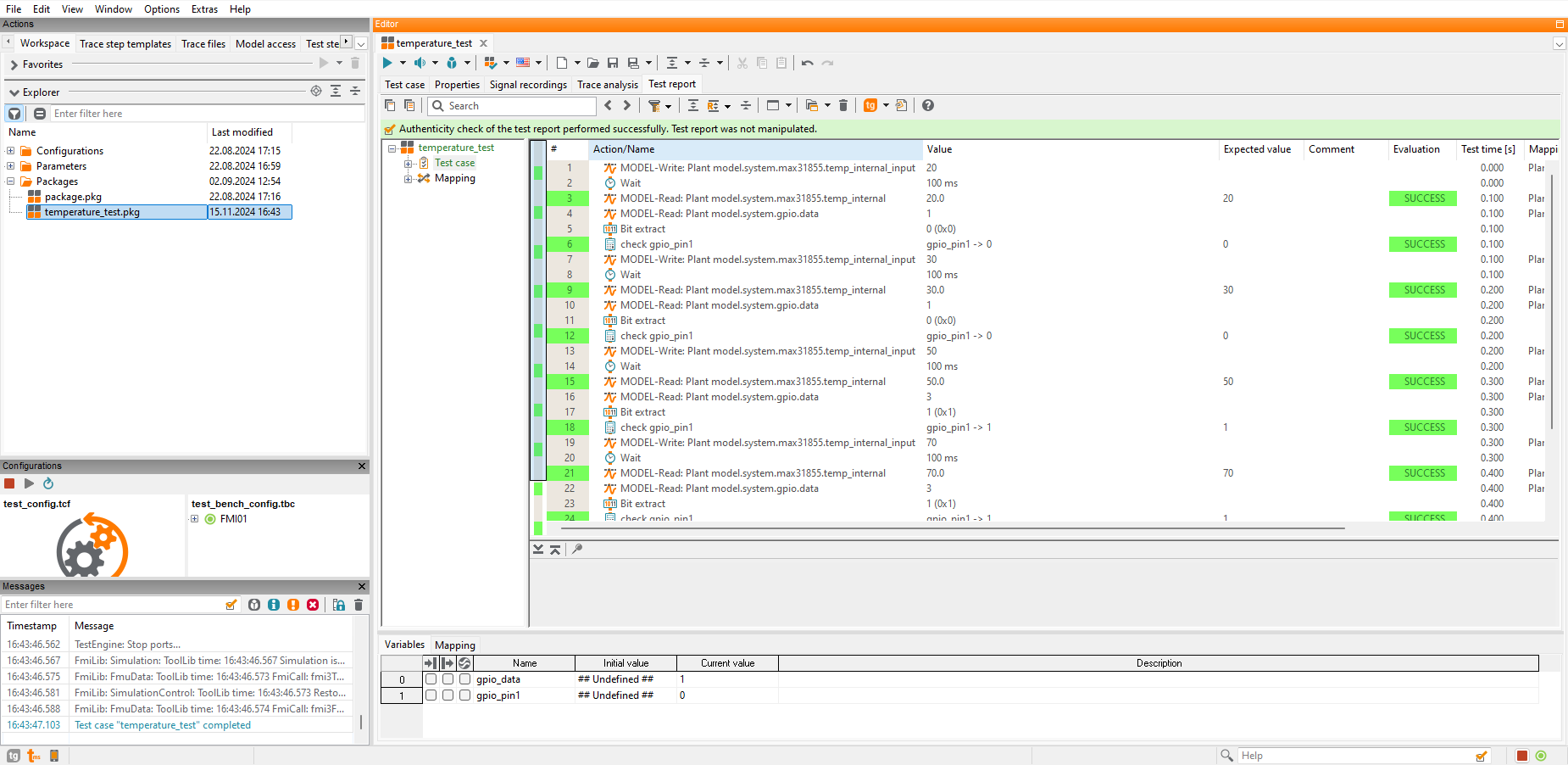}
  \caption{Successful test run in \ecutest.}
  \label{fig:tt_success}
\end{figure*}

\textbf{Environmental Impacts}: The \ac{ecu} interacts with its environment by measuring the ambient temperature and reacting to changes.
To add the simulation of changing environmental conditions and enable comprehensive testing with dynamic environmental scenarios, we package the \ac{vp} into an \ac{fmu} using our \ac{fmi} integration approach.

This allows for co-simulation with external tools that can provide varying temperature inputs.
The \ac{fmu} configuration for our scenario includes one input and one output as shown in \cref{fig:fmu_schema}.
The input is connected to the temperature property of the MAX31855 sensor model, and the output retrieves the \ac{gpio} pin value from the \ac{gpio} controller (see \cref{fig:vp}).
There are no changes in the \ac{vp} required to be connected to the \ac{fmu}.
It is sufficient to specify the hierarchical names of the properties for the temperature value and the \ac{gpio} pins in the model description (see \cref{sec:approach:fmu-creation}).

\subsection{Target-Software Verification}
\label{sec:case-study:target-sw-verification}

Software verification and testing play a crucial role during the design process of a system.
The earlier a bug is found, the less effort and cost it will take to fix it.
As described in \cref{sec:introduction}, if a bug is first found during product operation, the cost of fixing it can be up to \num{100} times higher than if it is found during software development \cite{boehm_software_1976}.
By integrating the \ac{vp} into an \ac{fmi} co-simulation using our approach, complex scenarios can be tested to find bugs as early as possible.
This leads to lower costs, a shorter time to market, and overall improved development efficiency.

A typical measure that is used to determine the percentage of tested code is code coverage \cite{ivankovic_code_2019}.
Traditional line coverage \cite{miller_systematic_1963} describes the number of executed lines of code divided by the total number of lines of code.
More advanced metrics like \ac{mcdc} \cite{kelly_j_hayhurst_practical_2001} are required for ISO-26262-based \ac{asil}~D certification \cite{iso-26262}.
Besides counting the executed lines of code, \ac{mcdc} adds additional metrics like counting if Boolean expressions have been evaluated as both, true and false.
The \ac{vp} can directly extract coverage metrics of the target software after execution without requiring instrumentation of the target software \cite{bosbach_nqc_2024}.

For extensive testing and validation, we utilize \textit{\ecutest} from \textit{tracetronic GmbH} \cite{tracetronic_gmbh_ecutest_2024}.
This tool allows defining complex test scenarios, specify environmental input data, and compare the system's responses against expected outcomes to test if the system fulfills its requirements.
\ecutest is commonly used for \ac{ecu} testing and verification.
It has native support for \ac{fmi}.

For easy setup and test configuration, \ecutest can be executed on Windows-based machines and controlled using a \ac{gui}.
The \ac{vp} is executed on a Linux cloud server.
\ecutest loads our adapter \ac{fmu} on the Windows machine.
The \ac{fmu} connects via a \ac{tcp}/\ac{ip} connection to the \ac{vsp} server that runs on a Linux machine in the same network.
For scaled testing, the whole setup can be executed on a single Linux machine using tracetronic's Linux version of \ecutest.
By combining our \ac{vp} \ac{fmu} with \ecutest, we can perform realistic and thorough testing of the Schmitt trigger application by simulating various temperature conditions.
An example of a successful test run is illustrated in \cref{fig:tt_success}.

Execution traces of the \ac{cpu} model can be used to generate coverage reports.
To showcase the usefulness of a coverage report, we turned on the code-coverage feature of the \ac{vp} to generate execution traces and converted them to a \textit{lcov}-based coverage report \cite{linux_test_project_ltp_2023}.

\begin{figure}[!t]
  \centering
  \begin{subfigure}[c]{.95\linewidth}
    \centering
    \fbox{\includegraphics[width=\linewidth]{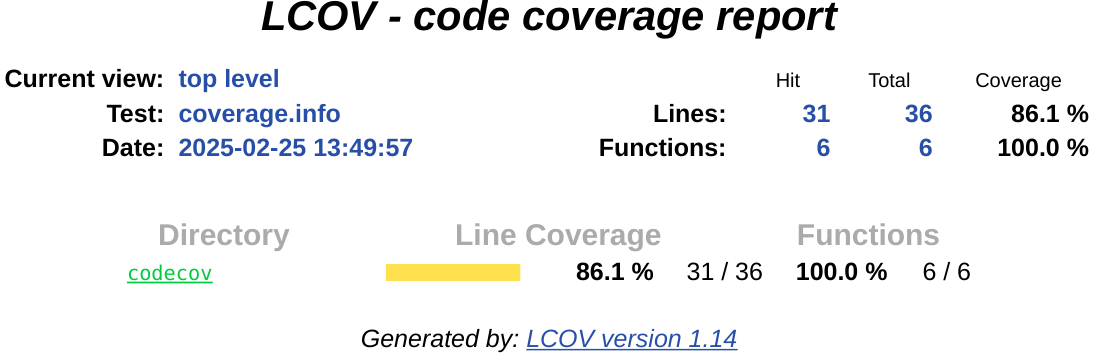}}
    \caption{Constant temperature (standalone \acs{vp}).}
    \vspace{1em}
    \label{fig:cov:const}
  \end{subfigure}\\

  \begin{subfigure}[c]{.95\linewidth}
    \centering
    \fbox{\includegraphics[width=\linewidth]{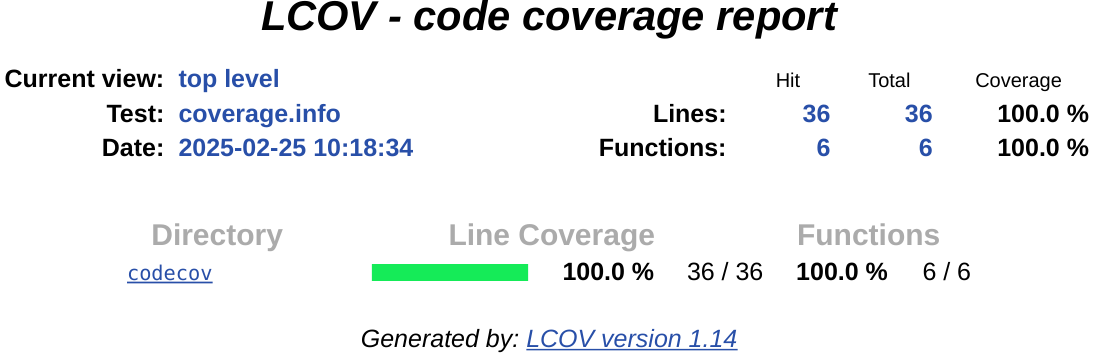}}
    \caption{Temperature controlled by \ecutest.}
    \vspace{1em}
    \label{fig:cov:test}
  \end{subfigure}
  \caption{Coverage results for different configurations.}
  \label{fig:cov}
  \vspace{-1em}
\end{figure}

\cref{fig:cov} visualizes coverage results for the presented Schmitt-trigger application.
\Cref{fig:cov:const} shows the report for the standalone \ac{vp} without the \ac{fmi} integration and \ecutest.
Since the temperature value is constant in this case, the algorithm never executes the path that sets the \ac{gpio} pin (see \cref{fig:alg}).
This leads to a line coverage of \SI{86.1}{\percent}.
When \ecutest is connected and test scenarios are executed, the line coverage is increased to \SI{100}{\percent}.

This simple example shows that the consideration of environmental influences is necessary for extensive testing.
The effectiveness of the applied testing is directly reflected by the increased code coverage.

A more realistic industrial-grade application is, e.g., the simulation of an engine control unit in a \ac{vp}.
The various sensor inputs (e.g., position, temperature, and veclocity sensors) can be read from \ac{fmi}.
Output-signal characteristics, such as the duty cycle and frequency of \ac{pwm} signals to control actuators like fuel injectors, ignition coils, and idle speed control valves, are forwarded to the importing tool for further processing.

\section{Design-Flow Integration}
\label{sec:design-flow-integration}

\begin{figure}[!t]
  \centering
  \input{tikz/design_flow.tikz}
  \vspace{0.5em}
  \caption{Incorporation of early target-software testing and validation in the design process.}
  \label{fig:design_flow}
\end{figure}
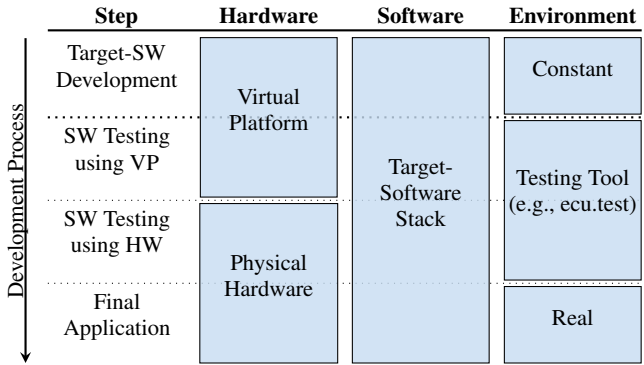

Based on our integration, we propose a design flow that can be used for early target software verification and creation of tests.
An overview is shown in \cref{fig:design_flow}.
At the beginning of the design process, a \ac{vp} can be created according to the specification of the system.
The target software can be developed and tested using this \ac{vp}.
Once the target software development is done, our \ac{fmi} integration can be used to integrate the \ac{vp} into a co-simulation and develop test scenarios.
Edge-cases can be tested to find bugs in the target software early.
The extracted coverage metrics guide verification and test engineers to cover all parts of the target software.

To fulfill standards like ISO~26262 \cite{iso-26262}, the satisfaction of the requirements needs to be tested.
Our \ac{fmu} can be used to design those tests using a \ac{vp}.
Once physical hardware is available, the hardware can directly be tested using the already developed tests.

In contrast to physical hardware, the virtual model can be scaled arbitrarily, distributed around the world within seconds, and easily integrated into \ac{cicd} flows.
\ac{cicd} allows for continuous and automatic testing during the development process to reduce costs by finding bugs earlier.

Once the physical hardware is available, the same test can be executed on the real hardware by replacing the adapter \ac{fmu} by an hardware-communication interface.
Without our co-simulation integration, the test would have needed to be created directly for the physical hardware, making it possible only later in the development process.

When the target software stack passed all tests on the \ac{vp} and the physical device, the application can be used in the real world application.

\section{Conclusion and Future Work}
In this paper, we have presented a novel approach to interface SystemC-based \acp{vp} with the \ac{fmi} standard for co-simulation.
By developing an adapter \ac{fmu} that communicates with the \ac{vp} via a \ac{tcp}/\ac{ip}-based connection, we enable the \ac{vp} to exchange data with external tools in a co-simulation environment.
Our method allows unmodified target software to run on the VP while interacting with realistic environmental inputs provided by other tools.

We demonstrated the effectiveness of our approach through a case study involving a Schmitt-trigger application running on the open-source \ac{avp64}.
By connecting the \ac{vp} to the \ac{fmi}-compatible \ecutest tool, we supplied dynamic temperature inputs and observed the corresponding behavior of the software, achieving comprehensive testing and improved code coverage.
This integration facilitates thorough software testing and verification by simulating environmental effects on the system, which is crucial for developing complex embedded systems.

While \ecutest is the leading testing tool in the automotive domain, and thereby a key candidate for evaluating our \ac{fmi} integration, \ac{fmi} opens the door to connect various other tools to \acp{vp} using our integration.
Users have the possibility co-simulate \acp{vp} with models created by other tools, e.g., to add realistic physical environmental models to sensors of the \ac{vp}.
Tool providers that support \ac{fmi} can directly benefit from our integration without the need of extending their tool.
Moreover, \ac{vp} developers can adopt a co-simulation approach to introduce physical environmental models when necessary, while retaining the flexibility to revert to a \ac{vp}-only simulation if those models are not required.

Since our method only relies on \ac{vcml}'s \ac{vsp} to establish the \ac{tcp}/\ac{ip} connection, it is not limited to a specific \ac{vp}.
There are no changes required for the \ac{vp} to work with \ac{fmi}, so other \acp{vp} can directly be integrated into an \ac{fmi} co-simulation.

We showcased how the approach can be integrated into the target-software design process to enable early test development and testing.
This speeds up the design process and reduces the number of errors in the product.

For future work, we plan to extend our approach to support more advanced communication between the \ac{vp} and other \acp{fmu} by adding support for the new \ac{fmi-ls-bus}.
This addition will allow connecting \acp{fmu} via virtual network interfaces like \ac{can} or Ethernet.

\printbibliography

\end{document}

%% file: tikz/schema.tikz
\pgfdeclarelayer{b1}
\pgfdeclarelayer{b2}
\pgfdeclarelayer{b3}
\pgfsetlayers{b3,b2,b1,main}

\begin{tikzpicture}
    \node[nodetype, fill=grun-25, text width=2.5cm, minimum height=1.2cm, text depth=0.7cm] (fmu) {\textbf{Adapter FMU}};
    \node[nodetype, fill=grun-50, text width=2.2cm, minimum height=0.5cm, anchor=south] at ([yshift=1mm]fmu.south) (vsp_client) {VSP Client};

    \node[nodetype, fill=rwth-25, text width=1.5cm, minimum height=1.2cm, anchor=east] at ([xshift=-1cm]fmu.west) (importing tool) {\textbf{Import Tool}};

    \node[nodetype, fill=rwth-25, text width=5.2cm, minimum height=2.1cm, text depth=1.5cm, align=left, anchor=north east] at ([yshift=-0.5cm]fmu.south east) (vp) {\textbf{\acl*{vp}}};
    \node[nodetype, fill=rwth-50, text width=2.2cm, minimum height=0.5cm, anchor=north] at ([yshift=-7mm]vsp_client.south) (vsp_server) {VSP Server};
    \node[nodetype, fill=rwth-50, text width=2.2cm, minimum height=0.5cm, anchor=south] at ([yshift=+1.5mm]vsp_server.center|-vp.south) (models) {Models};
    \node[nodetype, fill=rwth-50, text width=2cm, minimum height=0.5cm, anchor=east] at ([xshift=-3.5mm]models.west) (systemc) {SystemC Kernel};

    \phantom{\node[nodetype, inner sep=0pt, fit={(importing tool) (fmu)}] (process1_box) {};}
    \node[label, anchor=south, rotate=90] at ([xshift=-1mm]process1_box.west) (process1_label) {\textit{Process 1}};

    \phantom{\node[nodetype, inner sep=0pt, fit={(vp)}] (process2_box) {};}
    \node[label, anchor=south, rotate=90] at (process2_box.center-|process1_label.south) (process2_label) {\textit{Process 2}};

    \begin{pgfonlayer}{b1}
        \node[nodetype, densely dotted, fill=rwth-10, fit={(process1_box) (process1_label)}] (process1) {};
        \node[nodetype, densely dotted, fill=rwth-10, fit={(process2_box) (process2_label)}] (process2) {};
    \end{pgfonlayer}

    \tikzstyle{connector} = [draw=black,thick]
    \coordinate (conn_mid) at ($(fmu.west)!0.5!(importing tool.east)$);
    \draw[connector] ([xshift=-1mm,yshift=2mm]conn_mid) arc[start angle=90, end angle=270, radius=2mm];
    \draw[connector] ([xshift=1mm]conn_mid) circle(2mm);

    \draw[connector] ([xshift=-3mm]conn_mid) -- (importing tool.east);
    \draw[connector] ([xshift=+3mm]conn_mid) -- (fmu.west);

    \draw[<->,>=stealth,thick] (vsp_client) -- node[right] {\footnotesize\acs*{tcp}/\acs*{ip}} (vsp_server);

    \coordinate (time_start) at ([xshift=-2mm]vsp_server.south);
    \coordinate (time_end) at ([xshift=-7mm]systemc.north);
    \coordinate (time_mid) at ($(time_start)!0.5!(time_end)$);
    \draw[->,>=stealth,thick] (time_start) |- (time_mid) -| node[label, above, pos=0, yshift=1pt] {Simulation-Time Control} (time_end);

    \coordinate (prop_exchange) at ([xshift=0mm]vsp_server.south);
    \draw[<->,>=stealth,thick] (prop_exchange) -- ++(0, -2mm) -- node[label, pos=0.4, xshift=1mm, anchor=west] {\shortstack[l]{In/Out\\Exchange}} (prop_exchange|-models.north);

    \node[label, anchor=north] at ([xshift=10mm,yshift=-2mm]process1.north east) (legend_title) {\textbf{Legend}};

    \draw[connector] ([xshift=+2mm,yshift=-1mm]legend_title.south) arc[start angle=90, end angle=270, radius=2mm];
    \draw[connector] ([yshift=-3mm]legend_title.south) -- ++(-2mm,0mm);
    \node[label, anchor=north, text width=1.6cm, align=center] at ([yshift=-6mm]legend_title.south) (legend_provided_text) {Provided \acs*{fmi}-for-Co-Simulation};

    \draw[connector] ([xshift=1mm,yshift=-5mm]legend_provided_text.south) circle(2mm);
    \draw[connector] ([xshift=-1mm,yshift=-5mm]legend_provided_text.south) -- ++(-2mm,0mm);
    \node[label, anchor=north, text width=1.6cm, align=center] at ([yshift=-8mm]legend_provided_text.south) (legend_required_text) {Required \acs*{fmi}-for-Co-Simulation};

    \begin{pgfonlayer}{b1}
        \node[nodetype, fill=black-10, fit={(legend_title) (legend_required_text)}] {};
    \end{pgfonlayer}

\end{tikzpicture}

%% file: tikz/fmi_me.tikz
\pgfdeclarelayer{background}
\pgfsetlayers{background,main}

\begin{tikzpicture}
    \phantom{\node[nodetype, fill=rwth-50] (dummy_model) {Model};}
    \node[nodetype, fill=rwth-50, anchor=north] at([yshift=-0.1cm]dummy_model.south) (solver) {Solver};
    \node[label, anchor=south] at ([yshift=+0.1cm]dummy_model.north) (lbl_tool) {\textbf{Import Tool}};
    \begin{pgfonlayer}{background}
        \node[nodetype, fill=rwth-25, fit={(dummy_model) (solver) (lbl_tool)}] (tool) {};
    \end{pgfonlayer}

    \node[nodetype, fill=grun-50, anchor=west] at([xshift=1.1cm]dummy_model.east) (model) {Model};
    \phantom{\node[nodetype, fill=grun-25, anchor=center] at (solver.center-|model.center) (dummy_solver) {Solver};}
    \node[label, anchor=center] at (lbl_tool.center-|model.center) (lbl_fmu) {\textbf{FMU Lib}};
    \begin{pgfonlayer}{background}
        \node[nodetype, fill=grun-25, fit={(model) (dummy_solver) (lbl_fmu)}] (fmu) {};
    \end{pgfonlayer}

    \tikzstyle{connector} = [draw=black,thick]
    \coordinate (conn_mid) at ($(tool.east)!0.5!(fmu.west)$);
    \draw[connector] ([xshift=-1mm,yshift=2mm]conn_mid) arc[start angle=90, end angle=270, radius=2mm];
    \draw[connector] ([xshift=1mm]conn_mid) circle(2mm);

    \draw[connector] ([xshift=-3mm]conn_mid) -- (tool.east|-conn_mid);
    \draw[connector] ([xshift=+3mm]conn_mid) -- (fmu.west|-conn_mid);

\end{tikzpicture}

%% file: tikz/fmi_cs.tikz
\pgfdeclarelayer{background}
\pgfsetlayers{background,main}

\begin{tikzpicture}
    \phantom{\node[nodetype, fill=rwth-50, anchor=west] (dummy_model) {Model};}
    \phantom{\node[nodetype, fill=rwth-50, anchor=north] at([yshift=-0.1cm]dummy_model.south) (dummy_solver) {Solver};}
    \node[label, anchor=south] at ([yshift=+0.1cm]dummy_model.north) (lbl_tool) {\textbf{Import Tool}};
    \begin{pgfonlayer}{background}
        \node[nodetype, fill=rwth-25, fit={(dummy_model) (dummy_solver) (lbl_tool)}] (tool) {};
    \end{pgfonlayer}

    \node[nodetype, fill=grun-50, anchor=west] at([xshift=1.1cm]dummy_model.east) (model) {Model};
    \node[nodetype, fill=grun-50, anchor=center] at (dummy_solver.center-|model.center) (solver) {Solver};
    \node[label, anchor=center] at (lbl_tool.center-|model.center) (lbl_fmu) {\textbf{FMU Lib}};
    \begin{pgfonlayer}{background}
        \node[nodetype, fill=grun-25, fit={(model) (solver) (lbl_fmu)}] (fmu) {};
    \end{pgfonlayer}

    \tikzstyle{connector} = [draw=black,thick]
    \coordinate (conn_mid) at ($(tool.east)!0.5!(fmu.west)$);
    \draw[connector] ([xshift=-1mm,yshift=2mm]conn_mid) arc[start angle=90, end angle=270, radius=2mm];
    \draw[connector] ([xshift=1mm]conn_mid) circle(2mm);

    \draw[connector] ([xshift=-3mm]conn_mid) -- (tool.east|-conn_mid);
    \draw[connector] ([xshift=+3mm]conn_mid) -- (fmu.west|-conn_mid);

\end{tikzpicture}

%% file: tikz/fmi_cs_standalone.tikz
\pgfdeclarelayer{l1}
\pgfdeclarelayer{l0}
\pgfsetlayers{l1,l0,main}

\begin{tikzpicture}
    \phantom{\node[nodetype, fill=rwth-25, anchor=west] (dummy_model) {Model};}
    \phantom{\node[nodetype, fill=rwth-50, anchor=north] at ([yshift=-0.1cm]dummy_model.south) (dummy_solver) {Solver};}
    \node[label, anchor=south] at ([yshift=+0.1cm]dummy_model.north) (lbl_tool) {\textbf{Import Tool}};
    \begin{pgfonlayer}{l0}
        \node[nodetype, fill=rwth-25, fit={(dummy_model) (dummy_solver) (lbl_tool)}] (tool) {};
    \end{pgfonlayer}

    \node[nodetype, fill=grun-50, anchor=west] at([xshift=1.5cm]dummy_model.east) (model) {Model};
    \node[nodetype, fill=grun-50, anchor=center] at(model.center|-dummy_solver.center) (solver) {Solver};
    \node[label, anchor=center] at (lbl_tool.center-|model.center) (lbl_fmu) {\textbf{FMU Lib}};
    \begin{pgfonlayer}{l0}
        \node[nodetype, fill=grun-25, fit={(model) (solver) (lbl_fmu)}] (fmu) {};
    \end{pgfonlayer}

    \tikzstyle{connector} = [draw=black,thick]
    \coordinate (conn_mid) at ($(tool.east)!0.5!(fmu.west)$);
    \draw[connector] ([xshift=-1mm,yshift=2mm]conn_mid) arc[start angle=90, end angle=270, radius=2mm];
    \draw[connector] ([xshift=1mm]conn_mid) circle(2mm);

    \draw[connector] ([xshift=-3mm]conn_mid) -- (tool.east|-conn_mid);
    \draw[connector] ([xshift=+3mm]conn_mid) -- (fmu.west|-conn_mid);

    \node[label, anchor=south, rotate=90] at ([xshift=-1mm]tool.west) (p1_lbl) {\textit{Process 1}};
    \begin{pgfonlayer}{l1}
        \node[nodetype, densely dotted, fill=rwth-10, fit={(fmu) (tool) (p1_lbl)}] (p1) {};
    \end{pgfonlayer}
\end{tikzpicture}

%% file: tikz/fmi_cs_adapter.tikz
\pgfdeclarelayer{l1}
\pgfdeclarelayer{l0}
\pgfsetlayers{l1,l0,main}

\begin{tikzpicture}
    \phantom{\node[nodetype, fill=rwth-50, anchor=north] (dummy_solver) {Solver};}
    \node[label, anchor=south] at ([yshift=+0.1cm]dummy_solver.north) (lbl_tool) {\textbf{Import Tool}};
    \begin{pgfonlayer}{l0}
        \node[nodetype, fill=rwth-25, fit={(dummy_solver) (lbl_tool)}] (tool) {};
    \end{pgfonlayer}

    \node[nodetype, fill=grun-50, anchor=west] at([xshift=1.2cm]dummy_solver.east) (client) {Client};
    \node[label, anchor=center] at (lbl_tool.center-|client.center) (lbl_fmu) {\textbf{FMU Lib}};
    \begin{pgfonlayer}{l0}
        \node[nodetype, fill=grun-25, fit={(client) (lbl_fmu)}] (fmu) {};
    \end{pgfonlayer}

    \tikzstyle{connector} = [draw=black,thick]
    \coordinate (conn_mid) at ($(tool.east)!0.5!(fmu.west)$);
    \draw[connector] ([xshift=-1mm,yshift=2mm]conn_mid) arc[start angle=90, end angle=270, radius=2mm];
    \draw[connector] ([xshift=1mm]conn_mid) circle(2mm);

    \draw[connector] ([xshift=-3mm]conn_mid) -- (tool.east|-conn_mid);
    \draw[connector] ([xshift=+3mm]conn_mid) -- (fmu.west|-conn_mid);

    \node[label, anchor=south, rotate=90] at ([xshift=-1mm]tool.west) (p1_lbl) {\textit{Process 1}};
    \begin{pgfonlayer}{l1}
        \node[nodetype, densely dotted, fill=rwth-10, fit={(fmu) (tool) (p1_lbl)}] (p1) {};
    \end{pgfonlayer}

    \node[nodetype, fill=rwth-50, anchor=west] at([xshift=0.5cm]client.east) (server) {Server};
    \node[nodetype, fill=rwth-50, anchor=west] at([xshift=+0.1cm]server.east) (model) {Model};
    \node[nodetype, fill=rwth-50, anchor=north] at([yshift=-0.1cm]model.south) (solver) {Solver};
    \coordinate (sim_center) at ($(server.west)!0.5!(model.east)$);
    \node[label, anchor=center] at (lbl_tool.center-|sim_center) (lbl_sim) {\textbf{Simulation}};
    \begin{pgfonlayer}{l0}
        \node[nodetype, fill=rwth-25, fit={(server) (model) (solver) (lbl_sim)}] (sim) {};
    \end{pgfonlayer}

    \node[label, anchor=north, rotate=90] at ([xshift=+1mm]sim.east) (p2_lbl) {\textit{Process 2}};
    \begin{pgfonlayer}{l1}
        \node[nodetype, densely dotted, fill=rwth-10, fit={(sim) (lbl_sim) (p2_lbl) (p1_lbl.east-|p2_lbl.center)}] (p1) {};
    \end{pgfonlayer}

    \draw[<->, >=stealth, thick] (client) -- (server);
\end{tikzpicture}

%% file: tikz/creation.tikz
\pgfdeclarelayer{b1}
\pgfdeclarelayer{b2}
\pgfdeclarelayer{b3}
\pgfsetlayers{b3,b2,b1,main}

\begin{tikzpicture}

    \node[nodetype, fill=grun-25, text width=2.0cm] (mod_desc) {\shortstack{model\\Description.xml}};
    \node[nodetype, fill=grun-25, text width=2.0cm, anchor=north] at (mod_desc.south) (lib) {FMU Library};
    \node[nodetype, fill=grun-25, text width=2.0cm, anchor=north, densely dashed] at (lib.south) (bin) {(VP Binary)};
    \node[label, anchor=south] at ([yshift=0.1cm]mod_desc.north) (fmu) {\textbf{FMU}};

    \begin{pgfonlayer}{b1}
        \node[nodetype, fill=grun-10, fit={(mod_desc) (lib) (bin) (fmu)}] (fmu_box) {};
    \end{pgfonlayer}

    \node[nodetype, fill=rwth-25, text width=1.2cm, minimum height=1.0cm, anchor=south east] at ([xshift=-0.8cm]fmu_box.south west) (export) {Packager};

    \node[nodetype, fill=rwth-10, anchor=south east, minimum height=0.8cm, text width=2.0cm, densely dashed] at ([xshift=-0.8cm]export.south west) (vp) {VP Binary};
    \node[nodetype, fill=rwth-10, anchor=south east, minimum height=0.8cm, text width=2.0cm] at ([yshift=+0.1cm]vp.north east) (cfg) {\shortstack{model\\Description.xml}};
    \node[label, anchor=south] at ([yshift=+0.1cm]cfg.north) (input_lbl) {\textbf{Inputs}};

    \node[nodetype, fill=rwth-10, text width=1.2cm, anchor=south] at ([yshift=0.4cm]export.north) (fmu_src) {FMU Library};

    \draw[->, >=stealth, thick, shorten >=1mm, shorten <=1mm] (cfg.east) -- (export.west);
    \draw[->, >=stealth, thick, shorten >=1mm, shorten <=1mm, densely dashed] (vp.east) -- (vp.east-|export.west);
    \draw[->, >=stealth, thick, shorten >=1mm, shorten <=1mm] (export.east) -- (export.east-|fmu_box.west);
    \draw[->, >=stealth, thick, shorten >=1mm, shorten <=1mm] (fmu_src) -- (export);
\end{tikzpicture}

%% file: tikz/running_mode.tikz
\pgfdeclarelayer{b1}
\pgfdeclarelayer{b2}
\pgfdeclarelayer{b3}
\pgfsetlayers{b3,b2,b1,main}

\begin{tikzpicture}
    \node[nodetype, fill=rwth-25] (import_tool) {\textbf{Import Tool}};
    \node[nodetype, fill=grun-25, anchor=north] at ([yshift=-0.4cm]import_tool.south) (fmu) {\textbf{FMU Library}};

    \node[nodetype, fill=rwth-50, anchor=west] at ([xshift=3.0cm]$(import_tool.north)!0.5!(fmu.south)$) (vsp) {VSP Server};
    \node[label, anchor=south] at ([yshift=0.1cm]vsp.north) (vp) {\textbf{VP}};

    \begin{pgfonlayer}{b1}
        \node[nodetype, fill=rwth-25, fit={(vsp) (vp)}] (vp_box) {};
    \end{pgfonlayer}

    \phantom{\node[label, fit={(import_tool) (fmu) (vp_box.north-|import_tool.center) (vp_box.south-|import_tool.center)}] (p1_bb) {};}
    \phantom{\node[label, fit={(vsp) (vp) (vp_box) (import_tool.north-|vp.center) (fmu.south-|vp.center)}] (p2_bb) {};}

    \node[label, rotate=90, anchor=south] at ([xshift=-0.1cm]p1_bb.west) (p1_lbl) {\textit{Process 1}};
    \node[label, rotate=90, anchor=north] at ([xshift=+0.1cm]p2_bb.east) (p2_lbl) {\textit{Process 2}};

    \begin{pgfonlayer}{b2}
        \node[nodetype, densely dotted, fill=rwth-10, fit={(p1_bb) (p1_lbl)}] (p1_box) {};
        \node[nodetype, densely dotted, fill=rwth-10, fit={(p2_bb) (p2_lbl)}] (p2_box) {};
    \end{pgfonlayer}

    \draw[->,>=stealth,thick] (import_tool) -- node[label, right, xshift=0.1cm] {load} (fmu);
    \draw[->,>=stealth,thick] ([yshift=-0.1cm]fmu.east) -| node[label, below, pos=0.25, yshift=-1pt] {connect} (vsp.south);
    \draw[->,>=stealth,thick,dashed] ([yshift=+0.1cm]fmu.east) -| node[label, anchor=south east, pos = 0.4] {setup\textsuperscript{*}} ++(1.8cm,0.2cm) |- (p2_box.west);

    \node[label, anchor=north east] at ([yshift=-0.1cm]p2_box.south east) {\textsuperscript{*}if included in the FMU};
\end{tikzpicture}

%% file: tikz/call_seq.tikz
\pgfdeclarelayer{bg}
\pgfsetlayers{bg,main}

\begin{tikzpicture}
    \node[nodetype, fill=grun-25, text width=2.8cm, align=left] (instatiate) {\shortstack[l]{
            $\bullet$ Launch VP\textsuperscript{*}\\
            $\bullet$ Connect to VP
        }};
    \node[nodetype, fill=grun-25, text width=2.8cm, align=left, anchor=north west, yshift=-0.15cm, minimum height=0.5cm] at (instatiate.south west) (enterInit) {\shortstack[l]{
            $\bullet$ Simulate start time
        }};
    \node[nodetype, fill=grun-25, text width=2.8cm, align=left, anchor=north west, yshift=-0.15cm, minimum height=0.5cm] at (enterInit.south west) (getv) {\shortstack[l]{
            $\bullet$ Read value from VP
        }};
    \node[nodetype, fill=grun-25, text width=2.8cm, align=left, anchor=north west, yshift=-0.15cm, minimum height=0.5cm] at (getv.south west) (setv) {\shortstack[l]{
            $\bullet$ Write value to VP
        }};
    \node[nodetype, fill=grun-25, text width=2.8cm, align=left, anchor=north west, yshift=-0.15cm, minimum height=0.5cm] at (setv.south west) (step) {\shortstack[l]{
            $\bullet$ Simulate time step
        }};
    \node[nodetype, fill=grun-25, text width=2.8cm, align=left, anchor=north west, yshift=-0.15cm, minimum height=0.5cm] at (step.south west) (stop) {\shortstack[l]{
            $\bullet$ Stop VP
        }};
    \node[label, anchor=south, yshift=0.1cm] at (instatiate.north) (fmu) {\textbf{Adapter FMU}};

    \begin{pgfonlayer}{bg}
        \node[nodetype, fill=grun-10, fit={(instatiate) (enterInit) (getv) (setv) (step) (stop) (fmu)}] (fmu_box) {};
    \end{pgfonlayer}

    \node[label, anchor=east, xshift=-4.0cm] (importer) at (fmu_box.west) {\rotatebox{90}{\textbf{Import Tool}}};

    \begin{pgfonlayer}{bg}
        \node[nodetype, fill=rwth-10, fit={(importer) (importer.south|-stop.south) (importer.center|-fmu_box.north)}] (importer_box) {};
    \end{pgfonlayer}

    \draw[<-,>=stealth,thick] (instatiate.west) -- node[above, anchor=center] {\footnotesize\texttt{\shortstack{fmi3Instantiate\\CoSimulation}}} (importer_box.east|-instatiate.west);
    \draw[<-,>=stealth,thick] (enterInit.west) -- node[above, anchor=center] {\footnotesize\texttt{\shortstack{fmi3Enter\\InitializationMode}}} (importer_box.east|-enterInit.west);
    \draw[<-,>=stealth,thick] (getv.west) -- node[above, anchor=base, yshift=2pt] {\footnotesize\texttt{fmi3Get<datatype>}} (importer_box.east|-getv.west);
    \draw[<-,>=stealth,thick] (setv.west) -- node[above, anchor=base, yshift=2pt] {\footnotesize\texttt{fmi3Set<datatype>}} (importer_box.east|-setv.west);
    \draw[<-,>=stealth,thick] (step.west) -- node[above, anchor=base, yshift=2pt] {\footnotesize\texttt{fmi3DoStep}} (importer_box.east|-step.west);
    \draw[<-,>=stealth,thick] (stop.west) -- node[above, anchor=base, yshift=2pt] {\footnotesize\texttt{fmi3Terminate}} (importer_box.east|-stop.west);

    \draw[->,>=stealth,thick] ([xshift=-0.1cm]step.south east) |- ++(0.3cm,-0.08cm) |- ([xshift=-0.1cm]getv.north east);

    \node[label, anchor=north east] at ([yshift=-0.1cm]fmu_box.south east) {\textsuperscript{*}if included in the FMU};
\end{tikzpicture}

%% file: tikz/vp.tikz
\tikzstyle{nodetype} = [rectangle, rounded corners, minimum width=1.1cm, minimum height=.6cm, text centered, text width=1.1cm, draw=black, fill=rwth-25, font={\footnotesize}]
\tikzstyle{isock}    = [rectangle, minimum width=.15cm, minimum height=.15cm, draw=black, fill=black, inner sep=0pt, thick]
\tikzstyle{tsock}    = [isock, fill=white]
\tikzstyle{isocki}   = [isock, fill=rwth, draw=rwth]
\tikzstyle{tsocki}   = [isocki, fill=white]
\tikzstyle{isockspi} = [isock, fill=purple, draw=purple]
\tikzstyle{tsockspi} = [isockspi, fill=white]
\tikzstyle{isocksd}  = [isock, fill=orange, draw=orange]
\tikzstyle{tsocksd}  = [isocksd, fill=white]
\tikzstyle{lbl}      = [font={\tiny}, inner sep=1pt, text depth=0pt]
\tikzstyle{phantom}  = [inner sep=0pt, text depth=0pt, text height=0pt]
\tikzstyle{socklbl}  = [lbl]

\pgfdeclarelayer{background}
\pgfsetlayers{background,main}

\begin{tikzpicture}[node distance=.5cm and .5cm]

    \begin{scope}[name prefix=cpu-]
        \node[nodetype] (main) {CPU};
        \node[isock] at (main.south) (sock) {};
    \end{scope}

    \begin{scope}[name prefix=uart-]
        \node[nodetype, anchor=west] at ([xshift=1mm]cpu-main.east) (main) {UART};
        \node[tsock] at (main.south) (sock) {};
    \end{scope}

    \begin{scope}[name prefix=ram-]
        \node[nodetype, anchor=north] at ([yshift=-10mm]cpu-main.south) (main) {RAM};
        \node[tsock] at (main.north) (sock) {};
    \end{scope}

    \begin{scope}[name prefix=mmgpio-]
        \node[nodetype, anchor=west] at ([xshift=1mm]ram-main.east) (main) {GPIO};
        \node[tsock] at (main.north) (sock) {};
        \node[isocki] at ([xshift=-3mm]main.south) (socki) {};
    \end{scope}

    \begin{scope}[name prefix=spi-]
        \node[nodetype, anchor=west] at ([xshift=1mm]mmgpio-main.east) (main) {SPI};
        \node[tsock] at (main.north) (sock) {};
        \node[isockspi] at (main.south) (sockspi) {};
    \end{scope}

    \begin{scope}[name prefix=temp-]
        \node[nodetype, anchor=west] at ([xshift=1mm]spi-main.east) (main) {Sensor};
        \node[tsockspi] at (main.south) (sockspi) {};
    \end{scope}

    \begin{scope}[name prefix=sdhci-]
        \node[nodetype, anchor=west] at ([xshift=1mm]uart-main.east) (main) {SDHCI};
        \node[tsock, anchor=west] at ([xshift=0.5mm]main.south)(sock) {};
        \node[isock, anchor=east] at ([xshift=-0.5mm]main.south)(isock) {};
        \node[isocksd] at (main.east)(socksd) {};
    \end{scope}

    \phantom{\node[phantom, fit={(cpu-main) (uart-main) (sdhci-main)}] (components_upper) {};}
    \phantom{\node[phantom, fit={(ram-main) (mmgpio-main) (spi-main)}] (components_lower) {};}

    \begin{scope}[name prefix=bus-]
        \node[double arrow, draw, minimum height=4cm, minimum width=5mm, fill=rwth-25, font={\footnotesize}, anchor=center, inner sep=1pt,  double arrow head extend=0.1cm] at ($(components_upper.south)!0.5!(components_lower.north)$) (main) {System Bus};

        \node[tsock] at (main.north-|cpu-sock) (cpu){};
        \draw[thick] (cpu) -- (cpu-sock);

        \node[isock] at (main.north-|uart-sock) (uart){};
        \draw[thick] (uart) -- (uart-sock);

        \node[isock] at (main.north-|sdhci-sock) (sdhci){};
        \draw[thick] (sdhci) -- (sdhci-sock);
        \node[tsock] at (main.north-|sdhci-isock) (sdhcii){};
        \draw[thick] (sdhcii) -- (sdhci-isock);

        \node[isock] at (main.south-|ram-sock) (ram){};
        \draw[thick] (ram) -- (ram-sock);

        \node[isock] at (main.south-|mmgpio-sock) (rtc){};
        \draw[thick] (rtc) -- (mmgpio-sock);

        \node[isock] at (main.south-|spi-sock) (spi){};
        \draw[thick] (spi) -- (spi-sock);
    \end{scope}

    \begin{scope}[name prefix=sd-]
        \node[nodetype, anchor=center] at (bus-main.center-|temp-main.center) (main) {SD};
        \node[tsocksd] at (main.north)(socksd) {};
    \end{scope}
    \draw[thick,orange] (sdhci-socksd) -| +(2mm,-3mm) -| (sd-socksd);

    \phantom{\node[phantom, fit={(spi-main) (temp-main)}] (components_spi) {};}

    \begin{scope}[name prefix=spi-bus-]
        \node[double arrow, draw, minimum height=2.5cm, minimum width=4mm, fill=rwth-25, font={\footnotesize}, anchor=center, inner sep=1pt,  double arrow head extend=0.1cm] at ([yshift=-5mm]components_spi.south) (main) {SPI Bus};

        \node[tsockspi] at (main.north-|spi-sockspi) (spi){};
        \draw[thick, purple] (spi) -- (spi-sockspi);

        \node[isockspi] at (main.north-|temp-sockspi) (temp){};
        \draw[thick, purple] (temp) -- (temp-sockspi);
    \end{scope}

    \node[nodetype, anchor=west, text width=1.5cm] at ([xshift=4mm]temp-main.east) (vsp_server) {VSP Server};
    \draw[<->,>=stealth,thick] (vsp_server) -- (temp-main);
    \draw[<->,>=stealth,thick] (vsp_server.south) |- ([yshift=-0.2cm]spi-bus-main.south) -| (mmgpio-main);

    \begin{scope}[name prefix=vp-box-]
        \node[label, anchor=north east] at (sd-main.east|-sdhci-main.north) (lbl) {\textbf{\shortstack[r]{Virtual\\Platform}}};
        \begin{pgfonlayer}{background}
            \coordinate (top_left) at ([xshift=-1mm,yshift=1mm]cpu-main.north west);
            \coordinate (top_right) at ([xshift=1mm]lbl.east|-top_left);
            \coordinate (top_right2) at ([xshift=1mm,yshift=2mm]vsp_server.north east);
            \coordinate (bottom_left) at ([yshift=-2mm]spi-bus-main.after tip 1-|top_left);

            \draw[fill=rwth-10, rounded corners] (top_left)
                -- (top_right)
                |- (top_right2)
                |- (bottom_left)
                -- cycle;
        \end{pgfonlayer}
    \end{scope}

    \node[nodetype, anchor=south, fill=grun-50, text width=1.5cm] at ([yshift=5mm]vsp_server.north) (vsp_client) {VSP Client};
    \draw[<->,>=stealth,thick] (vsp_server) -- (vsp_client);

    \begin{scope}[name prefix=fmi-box-]
        \node[label, anchor=north] at (vsp_client.center|-vp-box-lbl.north) (lbl) {\textbf{Adapter FMU}};
        \begin{pgfonlayer}{background}
            \coordinate (top_left) at ([xshift=-1mm,yshift=1mm]lbl.north west);
            \coordinate (bottom_right) at ([xshift=1mm,yshift=-2mm]vsp_client.south east);
            \coordinate (top_right) at (top_left-|bottom_right);
            \coordinate (bottom_left) at (top_left|-bottom_right);

            \draw[fill=grun-25, rounded corners] (top_left)
                -- (top_right)
                -- (bottom_right)
                -- (bottom_left)
                -- cycle;
        \end{pgfonlayer}
    \end{scope}

    \begin{scope}[name prefix=legend-]
        \node[isock, anchor=south] at ([xshift=0cm,yshift=0.5cm]cpu-main.north) (isock) {};
        \node[lbl, anchor=west] at ([xshift=0.05cm]isock.east) (lbl_isock) {TLM Initiator};
        \node[tsock, anchor=west] at ([xshift=0.1cm]lbl_isock.east) (tsock) {};
        \node[lbl, anchor=west] at ([xshift=0.05cm]tsock.east) (lbl_tsock) {TLM Target};

        \node[isocki, anchor=west] at ([xshift=0.15cm]lbl_tsock.east) (isocki) {};
        \node[lbl, anchor=west] at ([xshift=0.05cm]isocki.east) (lbl_isocki) {IRQ Initiator};
        \node[tsocki, anchor=west] at ([xshift=0.1cm]lbl_isocki.east) (tsocki) {};
        \node[lbl, anchor=west] at ([xshift=0.05cm]tsocki.east) (lbl_tsocki) {IRQ Target};

        \node[isocksd, anchor=north] at ([yshift=-1mm]isock.south) (isocksd) {};
        \node[lbl, anchor=west] at ([xshift=0.05cm]isocksd.east) (lbl_isocksd) {SD Initiator};
        \node[tsocksd, anchor=center] at (isocksd.center-|tsock.center) (tsocksd) {};
        \node[lbl, anchor=west] at ([xshift=0.05cm]tsocksd.east) (lbl_tsocksd) {SD Target};

        \node[isockspi, anchor=center] at (isocki.center|-isocksd.center) (isockspi) {};
        \node[lbl, anchor=west] at ([xshift=0.05cm]isockspi.east) (lbl_isockspi) {SPI Initiator};
        \node[tsockspi, anchor=center] at (isockspi.center-|tsocki.center) (tsockspi) {};
        \node[lbl, anchor=west] at ([xshift=0.05cm]tsockspi.east) (lbl_tsockspi) {SPI Target};

        \begin{pgfonlayer}{background}
            \node[draw=black, fill=white, inner sep=1pt, fit={(isock) (lbl_tsockspi) (lbl_tsocki)}] {};
        \end{pgfonlayer}
    \end{scope}
\end{tikzpicture}

%% file: tikz/algorithm.tikz
\begin{tikzpicture}[node distance=0.3cm and 0.5cm]
    \node[startstop] (start) {\shortstack{\textbf{Start}\\\;\,$t_{up}$: Upper Threshold\\\;\;\,$t_{lo}$: Lower Threshold\\$p$: Polling Period}};
    \node[process, below=of start] (read_temp) {Read Temperature $t$};
    \node[decision, below=of read_temp] (upper_thr) {$t > t_{up}$};
    \node[decision, below=of upper_thr] (lower_thr) {$t \leq t_{lo}$};
    \node[process, right=of lower_thr] (clear_gpio) {Clear GPIO Pin};
    \node[process] at (clear_gpio|-upper_thr) (set_gpio) {Set GPIO Pin};
    \node[process, right=of set_gpio] (sleep) {Sleep $p$};

    \draw[->, >=stealth, thick] (start) -- (read_temp);
    \draw[->, >=stealth, thick] (read_temp) -- (upper_thr);
    \draw[->, >=stealth, thick] (upper_thr) -- node[label, right, inner sep=2pt, pos=0.3] {false} (lower_thr);
    \draw[->, >=stealth, thick] (upper_thr) -- node[label, above, inner sep=2pt, pos=0.3] {true} (set_gpio);
    \draw[->, >=stealth, thick] (lower_thr) -- node[label, above, inner sep=2pt, pos=0.3] {true} (clear_gpio);
    \draw[->, >=stealth, thick] (lower_thr.south) -- node[label, right, inner sep=2pt, pos=0.3] {false} ++(0cm,-0.3cm) -| ([xshift=0.4cm]sleep);
    \draw[->, >=stealth, thick] (clear_gpio.east) -| ([xshift=-0.4cm]sleep);
    \draw[->, >=stealth, thick] (set_gpio) -- (sleep);
    \draw[->, >=stealth, thick] (sleep) |- (read_temp);
\end{tikzpicture}

%% file: tikz/expected_behavior.tikz
\begin{tikzpicture}
    \begin{axis}[
        General,
        xmin=0,
        xmax=10.5,
        ylabel near ticks,
        yticklabel pos=right,
        ytick pos=right,
        ymajorgrids=false,
        axis x line=none,
        ylabel={GPIO Pin Value},
        ytick={0, 1},
        ymin=-0.1, ymax=1.2,
        width=7cm,
    ]
        \addplot[color=black, thick] coordinates {
            (0,0)
            (4,0)
            (4,1)
            (9,1)
            (9,0)
            (10,0)
        };
    \end{axis}

    \begin{axis}[
            General,
            xmin=0,
            xmax=10.5,
            xtick={0, 2, ..., 10},
            ytick={0, 10, ..., 60},
            xlabel={Simulation Time / \si{\second}},
            ylabel={Temperature / \si{\celsius}},
            ylabel style={color=rwth},
            yticklabel style={color=rwth},
            ymajorgrids=false,
            width=7cm,
        ]
        \addplot[color=bordeaux, densely dotted, thick] coordinates {
            (0,50)
            (10,50)
        };
        \node[anchor=west, color=bordeaux] at (axis cs: 10,50) {$t_{up}$};

        \addplot[color=grun, densely dotted, thick] coordinates {
            (0,40)
            (10,40)
        };
        \node[anchor=west, color=grun] at (axis cs: 10,40) {$t_{lo}$};

        \addplot[color=rwth, thick] coordinates {
            (0,10)
            (5,60)
            (7,60)
            (10,30)
        };
    \end{axis}
\end{tikzpicture}

%% file: tikz/fmu_schema.tikz
\pgfdeclarelayer{background}
\pgfsetlayers{background,main}

\begin{tikzpicture}
    \node[label] (fmu) {\textbf{FMU}};
    \phantom{\node[label, anchor=north] at ([yshift=-0.5cm]fmu.south) (fmu_bottom) {};}

    \begin{pgfonlayer}{background}
        \node[nodetype, fill=rwth-25, fit={(fmu) (fmu_bottom)}, inner sep=0.2cm, minimum width=1.8cm] (fmu_box) {};
    \end{pgfonlayer}

    \node[rectangle, anchor=north, draw=black, fill=white] at ([yshift=-0.2cm]fmu_box.west|-fmu.south) (in_temp_box) {};
    \node[rectangle, anchor=center, draw=black, fill=white] at (fmu_box.east|-in_temp_box.center) (out_gpio_box) {};

    \node[label, anchor=east] at ([xshift=-0.1cm]in_temp_box.west) (in_temp) {Temperature};
    \node[label, anchor=west] at ([xshift=+0.1cm]out_gpio_box.east) (out_gpio) {GPIO value};

    \draw  ([xshift=+0.1mm,yshift=-0.1mm]in_temp_box.north west)
        -- ([xshift=-0.2mm,yshift=+0.0mm]in_temp_box.east)
        -- ([xshift=+0.1mm,yshift=+0.1mm]in_temp_box.south west);

    \draw  ([xshift=+0.1mm,yshift=-0.1mm]out_gpio_box.north west)
        -- ([xshift=-0.2mm,yshift=+0.0mm]out_gpio_box.east)
        -- ([xshift=+0.1mm,yshift=+0.1mm]out_gpio_box.south west);
\end{tikzpicture}

%% file: tikz/design_flow.tikz
\tikzstyle{block} = [rectangle, minimum width=1.8, minimum height=.6cm, text centered, text width=1.6, draw=black, fill=rwth-25, fill opacity=0.8, text opacity=1, font={\footnotesize}, inner sep=0pt]

\pgfdeclarelayer{l0}
\pgfsetlayers{l0,main}

\begin{tikzpicture}[node distance=0.3cm and 0.5cm]
    \node[label, anchor=base] (lbl_step) {\textbf{Step}};
    \node[label, anchor=base] at ([xshift=2cm]lbl_step.base) (lbl_hw) {\textbf{Hardware}};
    \node[label, anchor=base] at ([xshift=2cm]lbl_hw.base) (lbl_sw) {\textbf{Software}};
    \node[label, anchor=base] at ([xshift=2cm]lbl_sw.base) (lbl_env) {\textbf{Environment}};

    \coordinate (top) at ([xshift=-1cm, yshift=-0.2cm]lbl_step.base);

    \begin{scope}[name prefix=step-dev-]
        \coordinate (top-left) at ([xshift=-0.9cm]lbl_step.center|-top);
        \coordinate (bottom-right) at ([xshift=1.8cm, yshift=-1.0cm]top-left);

        \node[block, fit={(top-left) (bottom-right)}, fill=none, draw=none] (box) {Target-SW Development};
    \end{scope}

    \begin{scope}[name prefix=step-test-]
        \coordinate (top-left) at ([xshift=-0.9cm, yshift=-0.1cm]lbl_step.center|-step-dev-bottom-right);
        \coordinate (bottom-right) at ([yshift=-1.0cm]top-left-|step-dev-box.east);

        \node[block, fit={(top-left) (bottom-right)}, fill=none, draw=none] (box) {SW Testing using VP};
    \end{scope}

    \begin{scope}[name prefix=step-test-hw-]
        \coordinate (top-left) at ([xshift=-0.9cm, yshift=-0.1cm]lbl_step.center|-step-test-bottom-right);
        \coordinate (bottom-right) at ([yshift=-1.0cm]top-left-|step-dev-box.east);

        \node[block, fit={(top-left) (bottom-right)}, fill=none, draw=none] (box) {SW Testing using HW};
    \end{scope}

    \begin{scope}[name prefix=step-app-]
        \coordinate (top-left) at ([xshift=-0.9cm, yshift=-0.1cm]lbl_step.center|-step-test-hw-bottom-right);
        \coordinate (bottom-right) at ([yshift=-1.0cm]top-left-|step-dev-box.east);

        \node[block, fit={(top-left) (bottom-right)}, fill=none, draw=none] (box) {Final Application};
    \end{scope}

    \coordinate (bottom) at (top|-step-app-box.south);

    \begin{scope}[name prefix=hw-vp-]
        \coordinate (top-left) at ([xshift=-0.9cm]lbl_hw.center|-step-dev-box.north);
        \coordinate (bottom-right) at ([xshift=1.8cm]top-left|-step-test-box.south);

        \node[block, fit={(top-left) (bottom-right)}] (box) {Virtual Platform};
    \end{scope}

    \begin{scope}[name prefix=hw-real-]
        \coordinate (top-left) at (hw-vp-box.west|-step-test-hw-box.north);
        \coordinate (bottom-right) at (hw-vp-box.east|-step-app-box.south);

        \node[block, fit={(top-left) (bottom-right)}] (box) {Physical Hardware};
    \end{scope}

    \begin{scope}[name prefix=hw-real-]
        \coordinate (top-left) at ([xshift=-0.9cm]lbl_sw.center|-step-dev-box.north);
        \coordinate (bottom-right) at ([xshift=1.8cm]top-left|-step-app-box.south);

        \node[block, fit={(top-left) (bottom-right)}] (box) {Target-Software Stack};
    \end{scope}

    \begin{scope}[name prefix=env-const-]
        \coordinate (top-left) at ([xshift=-0.9cm]lbl_env.center|-step-dev-box.north);
        \coordinate (bottom-right) at ([xshift=1.8cm]top-left|-step-dev-box.south);

        \node[block, fit={(top-left) (bottom-right)}] (box) {Constant};
    \end{scope}

    \begin{scope}[name prefix=env-testing-]
        \coordinate (top-left) at ([xshift=-0.9cm]lbl_env.center|-step-test-box.north);
        \coordinate (bottom-right) at ([xshift=1.8cm]top-left|-step-test-hw-box.south);

        \node[block, fit={(top-left) (bottom-right)}] (box) {Testing Tool (e.g., \ecutest)};
    \end{scope}

    \begin{scope}[name prefix=env-real-]
        \coordinate (top-left) at ([xshift=-0.9cm]lbl_env.center|-step-app-box.north);
        \coordinate (bottom-right) at ([xshift=1.8cm]top-left|-step-app-box.south);

        \node[block, fit={(top-left) (bottom-right)}] (box) {Real};
    \end{scope}

    \coordinate (right) at (top-|env-real-box.east);

    \begin{pgfonlayer}{l0}
        \coordinate (start) at ($(step-dev-box.south west)!0.5!(step-test-box.north west)$);
        \draw[dotted, thick] (start) -- (start-|right);

        \coordinate (start) at ($(step-test-box.south west)!0.5!(step-test-hw-box.north west)$);
        \draw[dotted] (start) -- (start-|right);

        \coordinate (start) at ($(step-test-hw-box.south west)!0.5!(step-app-box.north west)$);
        \draw[dotted] (start) -- (start-|right);

    \end{pgfonlayer}

    \draw[->, >=stealth, thick] ([xshift=-0.2cm]top) -- node[label, xshift=-0.1cm, rotate=90] {Development Process} ([xshift=-0.2cm]bottom);

    \coordinate (headline) at ($(lbl_step.base)!0.5!(step-dev-box.north)$);
    \draw[thick] (headline-|step-dev-box.west) -- (headline-|env-const-box.east);

\end{tikzpicture}